\documentclass[11pt,a4paper]{article}
\usepackage[T1]{fontenc}
\usepackage[utf8]{inputenc}
\usepackage{lmodern}
\usepackage[margin=26mm]{geometry}
\usepackage{amsmath,amssymb,amsthm}
\usepackage{booktabs,longtable,array}
\usepackage{xcolor}
\usepackage{graphicx}
\usepackage{enumitem}
\usepackage[round]{natbib}
\usepackage{algorithm}
\usepackage{algpseudocode}
\usepackage{microtype}
\usepackage{setspace}
\usepackage[hidelinks]{hyperref}
\graphicspath{{./}{Figures/}}
\hypersetup{
  pdftitle={Identification of Poisson processes through a one-dimensional projection},
  pdfauthor={Pablo A. Ferrari, Ricardo Fraiman, Sebastian Grynberg, Leonardo Moreno},
  pdfsubject={Point processes, Poisson process identification, Gaussian projections}
}
\allowdisplaybreaks

\theoremstyle{plain}
\newtheorem{theorem}{Theorem}
\newtheorem{proposition}[theorem]{Proposition}
\newtheorem{lemma}[theorem]{Lemma}
\newtheorem{corollary}[theorem]{Corollary}
\theoremstyle{definition}
\newtheorem{definition}{Definition}

\theoremstyle{remark}
\newtheorem{remark}{Remark}

\newcommand{\new}[1]{#1}

\newcommand{\R}{\mathbb{R}}
\newcommand{\N}{\mathbb{N}}
\newcommand{\Z}{\mathbb{Z}}
\newcommand{\PP}{\mathsf{P}}
\newcommand{\EE}{\mathsf{E}}
\newcommand{\Var}{\mathsf{V}}
\newcommand{\Cov}{\operatorname{Cov}}
\newcommand{\Leb}{\ell}
\newcommand{\1}{\mathbf 1}

\newcommand{\Poi}{\operatorname{Poisson}}

\newcommand{\Nu}{\mathcal N}
\newcommand{\Ucal}{\mathcal U}

\providecommand{\GT}{\mathrm{GT}}
\providecommand{\DPP}{\operatorname{DPP}}

\title{Identification of Poisson processes through a one-dimensional projection}
\author{Pablo A. Ferrari$^{1}$\thanks{Corresponding author: \texttt{pferrari@dm.uba.ar}} \and
Ricardo Fraiman$^{2,3}$ \and
Sebastian Grynberg$^{4}$ \and
Leonardo Moreno$^{5}$}
\date{}

\begin{document}
\maketitle
\begin{center}
\begin{minipage}{0.96\textwidth}
\small
$^{1}$ Departamento de Matem\'atica and IMAS, Facultad de Ciencias Exactas y Naturales, Universidad de Buenos Aires and CONICET, Argentina.\\
$^{2}$ Departamento de Matem\'atica y Ciencias, Universidad de San Andr\'es, Argentina.\\
$^{3}$ PEDECIBA, Matem\'atica, Uruguay.\\
$^{4}$ Departamento de Matem\'atica, Facultad de Ingenier\'ia, Universidad de Buenos Aires, Argentina.\\
$^{5}$ IESTA, Departamento de M\'etodos Cuantitativos, FCEA, Universidad de la Rep\'ublica, Uruguay.
\end{minipage}
\end{center}

\begin{abstract}
We study the identification of stationary Poisson processes in Euclidean space through a single one-dimensional projection. Given a point process and a straight line, each point is retained independently with a probability given by a Gaussian function of its distance to the line; the retained points are projected onto the line. 	
		If the process is Poisson, the projection is again Poisson.
		What a single projection reveals depends on the sign of the second factorial cumulant, which is also the sign of the two-point correlations. Under a Palm domination condition, which holds for positively associated processes, the cumulant is nonnegative. Using a Chen--Stein argument we show that when the cumulant of the projection vanishes, the process is Poisson. Consequently,  a single projection identifies non-Poissonianity. The same result holds under negative association. 
		When the cumulant changes sign, a random projection detects a nonzero cumulant with probability one. 
		We propose a test  whose rejection region is computed by Monte Carlo. The test, together with a spacings-based variant for repulsive alternatives, rejects determinantal, shot-noise and other classical  processes. When applied to lunar impact craters, the test rejects the Poisson hypothesis for craters with diameter below 30~km and does not reject it for craters above 45~km.
\end{abstract}

\section{Introduction}
	\label{sec:introduccion}
	
	Let $S$ be a simple stationary point process on $\R^d$. The question motivating this work is the following: \emph{can the non-Poissonianity of the process be identified through a single one-dimensional projection, and does that projection need to be chosen at random?} The answer to the first question is affirmative, at the level of distributions, for a broad class of processes satisfying a Palm domination condition; for these processes the answer to the second one is negative: any projection, fixed in advance, will do.
	
	The construction consists in choosing a line, thinning the point process with a Gaussian probability depending on the transversal distance of each point to the line, and projecting the retained points onto the longitudinal coordinate.  Independent thinning preserves the Poisson property. The Gaussian retention probability plays two roles: it is strictly positive, so that every pair of points of the process is seen by the projection, and its Gaussian form makes the projected second-order structure depend real analytically on the direction and the scale of the projection.
	
	In what follows, $S$ is a simple point process on $\R^d$ with distribution $\PP $. The process $S$ is Poisson when its counts over disjoint regions are independent and Poisson distributed. At the level of correlations, the deviation of a generic $S$ from the Poisson model is encoded by $c_2$, the second-order factorial cumulant density of $S$, whose vanishing is equivalent to the absence of second-order correlation between its points. The second factorial cumulant of a smoothed projection of $S$ is an explicit Gaussian average of $c_2$ along lines (Lemma~\ref{lem:cumulante-proyectada}).
	
	For a simple stationary point process $S$ on $\R^d$ with distribution $\PP $ let $\PP ^0$ be its Palm version. Heuristically, $\PP ^0$ describes the distribution of $S$ as seen from a typical point of the process, or simply the measure $\PP $ conditioned on the presence of a point at the origin,
	$
	\PP ^0(\cdot)=\PP (\cdot \mid 0\in S). 
	$
	The precise definition of $P^0$ is given in Section~\ref{sec:identificacion}. A process $S$ is stochastically dominated by $S'$ if $\EE f(S)\le \EE f(S')$ for all non decreasing $f$; in this case we denote $S\preceq S'$. 
	
	Assume $S$ is a stationary process satisfying $S\cup \{0\}\preceq S^0$. Denote by $Q_u$ the distribution of the smoothed projection of $S$ over a line, where the parameter $u\in\Ucal$ codifies the line and the standard deviation of the Gaussian probability, and by $Q_u^\ast$ the corresponding distribution when $S$ is replaced by a Poisson process with the same intensity. The identification mechanism can be summarized by the chain of implications
	$
	Q_u=Q_u^\ast\ \text{for one } u\in\Ucal
	\ \Longrightarrow\
	c_2^u=0
	\ \Longrightarrow\
	c_2=0
	\ \Longrightarrow\
	S\text{ is Poisson}
	$,
	where $c_2^u$ is the second factorial cumulant density of the projection. The first step is a property of Poisson processes, whose factorial cumulants of order at least two vanish. The second one uses that Palm domination forces $c_2\ge0$: since $c_2^u$ is an average of $c_2$ with a strictly positive Gaussian weight, it can only vanish if $c_2$ does. The last one is an application of Palm theory. If $S\cup\{0\}\preceq S^0$, then the second-order factorization, equivalent to $c_2=0$, allows a Chen--Stein argument based on void probabilities to conclude that the full process is Poisson. The same chain holds under the reverse domination $S\cup\{0\}\succeq S^0$, satisfied by negatively associated processes, in which case $c_2\le0$.
	
	By contraposition, if $S$ is not Poisson, then	$
	Q_u\neq Q_u^\ast$ for every  $u\in\Ucal,
	$ so that any single projection, fixed in advance, identifies non-Poissonianity. Without Palm domination in either direction $c_2$ may change sign, and the second step can fail for prescribed projections (Remark~\ref{rem:contraejemplo}). Real analyticity in the projection parameters still shows that, if $c_2\neq0$,
	$
	\Leb\{u\in\Ucal:Q_u=Q_u^\ast\}=0,
	$
	so that, if $U$ has an absolutely continuous distribution on the parameter space,
	$
	\PP\{Q_U=Q_U^\ast\}=0;
	$
	in this case the projection has to be drawn at random.
	
	The result is a statement about distributions, not about data. It guarantees that a non-Poisson process is not seen as Poisson by the projections described above. It says nothing about the distance between the law of $S$ and Poisson, for some distance between probabilities, which may be arbitrarily small. Estimating the distance from pure randomness, using finitely many points observed in a bounded window, remains a question of power, it depends on the intensity of the process, the size of the window, and the strength of the second-order structure. We do not discuss this issue in the paper.
	
	The result connects with the characterizations of the Poisson process through thinning \citep{AssuncaoFerrari2007}, with Palm and Chen--Stein techniques \citep{ChenXia2004,BobrowskiSchulteYogeshwaran2022,last-penrose}, and with identification results based on random projections of Cramér--Wold type \citep{CuestaFraimanRansford2006,CuestaFraimanRansford2007,FraimanMorenoRansford2024,FraimanMorenoRansford2025}. The main contribution is not a uniform power dominance over existing spatial tests, but a generic identification guarantee.
	
	The paper focuses on the Gaussian projection, the second factorial cumulant, Palm domination, and the one-projection theorem, and includes the statistical implementation of the procedure.
	
	\new{The proofs of Proposition~\ref{prop:chen-stein} and Lemmas~\ref{lem:fkg-palm}, \ref{lem:cumulante-proyectada} and~\ref{lem:signo} are given in Appendix~\ref{app:proofs}. Appendix~\ref{app:simulations} reports a simulation study that delimits the kinds of departure from the Poisson hypothesis to which the procedure is sensitive and those to which it is not: it compares the proposed test with three benchmark procedures under clustered and heterogeneous alternatives, and treats the repulsive case separately.}
	Lastly, impact craters in the lunar surfaces are analyzed by studying Poissonianity as a function of the diameter of the craters.
	
	\section{Smoothed Gaussian projections}
	\label{sec:proyecciones}
	
	We work in $\R^2$ to simplify the notation, although the construction extends to $\R^d$. For $\theta\in(0,\pi)$, define
	$
	e_\theta=(\cos\theta,\sin\theta),
	\qquad
	n_\theta=(-\sin\theta,\cos\theta).
	$
	The undirected line with normal $n_\theta$ and displacement $p\in\R$ is written as
	$	L_{\theta,p}=\{x\in\R^2:\langle x,n_\theta\rangle=p\}.
	$
	For fixed $\sigma>0$, let
	$
	w_\sigma(r)=\exp\left(-\frac{r^2}{2\sigma^2}\right),
	$
	\begin{align}
		\mbox{For} \ u=(\theta,p,\sigma) \ \mbox{denote} \
		\pi_u(x)=\langle x,e_\theta\rangle,
		\qquad
		q_u(x)=w_\sigma(\langle x,n_\theta\rangle-p),
	\end{align}
	that is, $\pi_u(x)$ is the projected coordinate of $x$ on the line $L_{\theta,p}$ and $q_u(x)$ is the Gaussian weight of $x$ associated to $u$.
	\begin{definition}[Smoothed Gaussian projection]
		\label{def:smoothed-projection}
		Let $u=(\theta,p,\sigma)$.		Conditionally on $S$, for each point $x\in S$ associate independent random variables
		$
		B_u(x)\sim
		\operatorname{Bernoulli}(q_u(x)).
		$
		The thinned Gaussian projection of $S$ associated to $u$ is defined as the point process on $\R$
		\begin{align}
			\Pi_u(S)
			:= \{ \pi_u(x)  : x\in S, B_u(x) =1\}.
		\end{align}
	\end{definition}
	
	\begin{proposition}[Behaviour under the Poisson
		\new{hypothesis}]
		\label{prop:poisson-projection}
		If $S$ is a stationary Poisson process on $\R^2$ with intensity $\lambda>0$, then, for every $u=(\theta,p,\sigma)$, $\Pi_u(S)$ is a stationary Poisson process on $\R$ with intensity
		\begin{align}
			\label{19}
			\lambda_\sigma
			=
			\lambda\int_\R w_\sigma(r)\,dr
			=
			\lambda\sqrt{2\pi}\,\sigma.
		\end{align}
	\end{proposition}
	
	This is immediate from the Colouring and Mapping Theorems for Poisson
	processes \citep[\S2.3, \S5.1]{kingman1993}: independent thinning multiplies
	the intensity by the retention probability, and the image under a measurable
	map is again Poisson with the push-forward intensity measure; integrating
	$\lambda w_\sigma$ over the transversal coordinate gives $\lambda_\sigma$,
	independently of $p$ by translation invariance.

	Let $S$ be a simple stationary point process on $\R^d$ with intensity
	$\lambda$. We say that $S$ has a second-order (product) density $\lambda(x,y)$ if
	\begin{equation}\label{eq:segundo-orden}
		\EE\big(|S\cap\Lambda_1|\,|S\cap\Lambda_2|\big)
		=\int_{\Lambda_1}\int_{\Lambda_2}\lambda(x,y)\,dx\,dy,
		\qquad \Lambda_1\cap\Lambda_2=\emptyset,
	\end{equation}
	for bounded Borel sets $\Lambda_1,\Lambda_2$
	\citep[][Ch.~4]{DaleyVereJones2003,last-penrose}. Its second
	factorial cumulant density is
	\begin{equation}\label{eq:c2}
		c_2(x,y)=\lambda(x,y)-\lambda^2=\kappa(y-x),
	\end{equation}
	where the last identity follows from stationarity, and $\kappa$ is even
	because $c_2$ is symmetric. For a stationary Poisson process
	$\lambda(x,y)=\lambda^2$, so that $c_2=0$; in general, $c_2$ measures the
	second-order correlation between the points of $S$.

	\section{Palm domination and Poisson characterization}
	\label{sec:identificacion}

	In this section $S$ denotes a simple stationary point process on $\R^d$ with intensity $\lambda\in(0,\infty)$, defined on the space $\mathbf{N}=\mathbf{N}(\R^d)$ of locally finite counting measures, with distribution $\PP $. The \emph{Palm distribution} of $S$ is the distribution $\PP^0$ of a process $S^0$ in $\mathbf{N}$ determined by the Campbell--Mecke formula {\citep[see][Ch.~9]{Mecke1967,last-penrose}}, for every measurable function $f:\R^d\times\mathbf{N}\to[0,\infty)$,
	\begin{align}
		\EE \sum_{x\in S} f(x,\vartheta_x S)
		\;=\; \lambda \int_{\R^d}
		\EE f(x,S^0)\,dx , \label{7}
	\end{align}
	where $\vartheta_xS=S-x$, that is $S$ as seen from $x$. The Palm version $S^0$  satisfies $0\in S^0$ almost surely and is interpreted as the process seen from a typical point of~$S$.
	
	Let $S$ and $S'$ be point processes with distribution $\PP$ and $\PP'$, respectively.  We say that $S$ is stochastically dominated by $S'$, and write $S\preceq S'$, when $\EE f(S)\leq \EE f(S')$ for every increasing local function $f$. This is equivalent to the existence of a coupling $(\hat S,\hat S')$, with marginals $\PP$ and $\PP'$ satisfying
	$\hat S\subset \hat S'$ \citep[see][]{thorisson,KamaeKrengelOBrien1977}.
	\begin{proposition}[Chen--Stein criterion]
		\label{prop:chen-stein}
		Let $S$ be a stationary point process on $\R^d$ with intensity $\lambda>0$. Assume
		\begin{gather}
			\text{either } \;     S\cup\{0\}\preceq S^0  \;\text{ or }  \;  S\cup\{0\}\succeq S^0   , \label{14}
			\\[1mm]
			\lambda(x,y)=\lambda^2,\qquad x\neq y.  \label{6}
		\end{gather}
		Then $S$ is a stationary Poisson process of intensity $\lambda$.
	\end{proposition}
	\new{The proof is given in Appendix~\ref{app:proofs}; it uses the following notation and basic properties.} 
	Let $S$ be a stationary process with intensity $\lambda$, and $\Lambda\subset\R^d$ be a compact set. Denote by $S^x\overset{d}{=} S^0+x$, the Palm version of $S$ conditioned on a point at $x$. Define
	$
	W=|S\cap\Lambda|,\qquad		W_x=|S^x\setminus\{x\}\cap\Lambda|$.
	
	We now compute the variances associated to $S$ in function of $c_2$.
	\begin{lemma}[Moments of $W$ and $W_x$]
		\label{lem:varianza}
		Let $S$ be a simple stationary point process on $\R^d$ with intensity $\lambda$ and second-order product density $\lambda(x,y)$, and let $\Lambda\subset\R^d$ be compact. Then, denoting $\Var$ the variance, 
		\begin{equation}
			\label{eq:varianza}
			EW= \lambda\,\Leb(\Lambda),\qquad \Var W
			=
			\EE W
			+
			\int_\Lambda\int_\Lambda c_2(x,y)\,dx\,dy,
		\end{equation}
		where $c_2(x,y)=\lambda(x,y)-\lambda^2$.
		If $\lambda(x,y)$ satisfies \eqref{6}, then
		\begin{align}
			\label{20}
			\Var W 	= 	\EE W 	= \EE W_x. 
		\end{align}
	\end{lemma}
	
	\begin{proof}
		Expression \eqref{eq:varianza} follows from Proposition 8.8 in \citet{last-penrose}.
		By the Campbell--Palm formula \citep[Ch.~9]{last-penrose}, \eqref{20} follows from
		$
		\EE W_x=\frac{1}{\lambda}\int_\Lambda \lambda(x,y)\,dy = \frac{\lambda^2}{\lambda} \ell(\Lambda)= \EE W. \qedhere
		$.
	\end{proof}
	
	\begin{lemma}
		\label{13}
		Let $S$ be a stationary process with intensity $\lambda$. Let $g:\N_0\to\R$ be a bounded function. Then,
		\begin{align}
			\label{9}
			\EE (Wg(W))= \lambda \int_\Lambda dx \,\EE g(W_x+1)
		\end{align}
	\end{lemma}
	\begin{proof}
		For bounded $g$, write
		\begin{align}
			\EE (Wg(W))
			&= \EE\sum_{x\in S}\, 1\{x\in\Lambda\}\,g(|S\cap\Lambda|)
			\label{8}\\
			&= \lambda\EE \int dx \,
			1\{x\in\Lambda\}\, g(|S^0\cap\Lambda-x|)
			\label{10}\\
			&= \lambda \int dx\, 1\{x\in\Lambda\}\,\EE g(|S^x\cap\Lambda|)
			\label{11}\\
			&= \lambda \int_\Lambda dx \,\EE g(W_x+1),\label{12}
		\end{align}
		where \eqref{10} is the Campbell identity (9.15) in \citet{last-penrose} for  $f(x,S) := 1\{x\in\Lambda\}\,g(|S\cap\Lambda|)$.
	\end{proof}

	The Palm domination conditions admit a verifiable criterion in terms of positive or negative association. 
	\begin{definition}[Positive and negative association]
		\label{def:fkg}
		A point process $S$ is positive associated if
		\begin{align}
			\EE(f(S)g(S))\geq \EE f(S)\,\EE g(S),\label{18}
		\end{align}
		for any increasing local functions $f$ and $g$ with well defined expectations. The positive association \eqref{18} is also called FKG property, see \citet{MR341553}.
		
		A point process $S$ is negative associated if
		\begin{align}
			\EE(f(S)g(S))\leq \EE f(S)\, \EE g(S),\label{21}
		\end{align}
		for increasing local functions $f$ and $g$ with well defined expectations and disjoint supports, that is, such that $f(S)=f(S\cap A)$ and $g(S)=g(S\cap B)$ for some disjoint bounded Borel sets $A$ and $B$.
	\end{definition}

	\begin{lemma}
		\label{lem:fkg-palm}
		Let $S$ be a simple stationary point process with finite positive intensity. 
		\begin{align}
			\text{If  $S$ is positive associated, then }  	&S\cup\{0\}\preceq S^0;\label{22}\\
			\text{if  $S$ is {negative} associated, then }  	&S\cup\{0\}\succeq S^0.\label{23}
		\end{align}
	\end{lemma}
	\new{The proof is given in Appendix~\ref{app:proofs}.}

	\section{Projected second factorial cumulant}\label{sec:segunda-cumulante}

	Let $S$ be a simple stationary point process on $\R^2$ with intensity
	$\lambda$ and second-order density $\lambda(x,y)$ as in
	\eqref{eq:segundo-orden}, and let $c_2$ and $\kappa$ be as in \eqref{eq:c2},
	with $\kappa$ locally integrable. The parameter space is
	$\Ucal=(0,\pi)\times\R\times\Sigma$, where $\Sigma\subset(0,\infty)$ is an
	open interval of scales, endowed with the Lebesgue measure $\ell$. For
	$u=(\theta,p,\sigma)\in\Ucal$ we write $Q_u$ and $Q_u^*$ for the laws of
	$\Pi_u(S)$ and $\Pi_u(S^*)$, where $S^*$ is a stationary Poisson process
	with constant intensity $\lambda$; by stationarity, $Q_u$ does not depend on $p$. Denote
	\begin{align}
		c_2^{u}(t,s)=\kappa_u(s-t),
	\end{align}
	the second factorial cumulant density of $\Pi_u(S)$.
	
	Since the projection of a Poisson process is also a Poisson process, it has vanishing second factorial cumulant. This section studies the reciprocal implication, namely, establish conditions on $S$ such that  $c^u_2=0$ implies $c_2=0$. The conditions depend on
	the sign of $c_2$. If $c_2$ is not zero and has constant sign, which is the
	case under Palm domination, then $c^u_2$ is also not zero with constant sign, and every single projection $u$ detects it, so that one can fix the projection in advance. If $c_2$ is not zero and changes sign, {then} $c^u_2$ is also not zero and changes sign for almost every projection $u$, but a prescribed finite family of
	projections may fail; in this case the projection has to be drawn at random.
	Finally, if $c_2=0$, the second-order structure carries no information;
	under Palm domination this case corresponds to the Poisson process
	(Proposition~\ref{prop:chen-stein}).
	
	We start by computing the projected
	cumulant.
	\begin{lemma}[Projected cumulant]\label{lem:cumulante-proyectada}
		Assume that $\kappa$ has constant sign or that
		\begin{equation}\label{eq:integrabilidad}
			\int_{\R^2}|\kappa(z)|\,e^{-\varepsilon|z|^2}\,dz<\infty
			\qquad\text{for every }\varepsilon>0 .
		\end{equation}
		Then
		\begin{equation}\label{eq:cumulante-proyectada}
			\kappa_u(\tau)=\sqrt{\pi}\,\sigma\int_{\R}
			\kappa\big(\tau e_\theta+v\,n_\theta\big)\,
			e^{-v^2/4\sigma^2}\,dv .
		\end{equation}
	\end{lemma}
	\new{The proof is given in Appendix~\ref{app:proofs}.}
	In particular, $\kappa_u$ does not depend on $p$, and if $\Pi_u(S)$ is a
	Poisson process, that is  $Q_u=Q_u^*$, then $\kappa_u=0$ almost
	everywhere.

	Formula \eqref{eq:cumulante-proyectada} shows that $\kappa_u$ is an average
	of $\kappa$ along lines, with a strictly positive weight. Whether this
	average can vanish without $\kappa$ vanishing depends on the sign of
	$\kappa$, and the sign is fixed by Palm domination.
	
	\paragraph{The second cumulant $c_2$ has a fixed sign}
	
	\begin{lemma}[Palm domination forces a fixed sign]\label{lem:signo}
		If $S\cup\{0\}\preceq S^0$, then $\kappa\ge 0$ almost everywhere.
		Symmetrically, if $S^0\preceq S\cup\{0\}$, then $\kappa\le 0$ almost
		everywhere.
	\end{lemma}
	
	\new{The proof is given in Appendix~\ref{app:proofs}.}
	
	With a fixed sign there are no cancellations in
	\eqref{eq:cumulante-proyectada}, and a single projection, chosen in advance,
	already detects $c_2\neq 0$.
	
	\begin{proposition}[Fixed sign: any given projection suffices]
		\label{prop:una-proyeccion}
		Assume that $\kappa\ge 0$ (or $\kappa\le 0$) almost everywhere and that
		$\kappa$ is not almost everywhere zero. Then, for every $u\in\Ucal$, $c^u_2$ is not almost everywhere zero. In particular no projection $\Pi_u(S)$,
		$u\in\Ucal$, is a Poisson process, $Q_u\neq Q_u^*$ for
		every $u\in\Ucal$.
	\end{proposition}
	
	\begin{proof}
		If $\Pi_u(S)$ is Poisson, then $\kappa_u=0$ almost everywhere by
		Lemma~\ref{lem:cumulante-proyectada}. Since the integrand in
		\eqref{eq:cumulante-proyectada} has constant sign and the weight is strictly
		positive, $\kappa(\tau e_\theta+vn_\theta)=0$ for almost every $(\tau,v)$,
		that is, $\kappa=0$ almost everywhere, a contradiction.
	\end{proof}

	The set of projections that fail to detect $S$ is therefore empty, not
	merely negligible, and neither analyticity nor a random choice of $u$ is
	needed.
	
	\paragraph{The second cumulant $c_2$ changes sign}
	
	When $c_2$ changes sign, cancellations in \eqref{eq:cumulante-proyectada}
	may occur, but only on a negligible set of projections, because the
	Gaussian kernel makes $\kappa_u$ depend analytically on $(\theta,\sigma)$.
	To see this, test $\kappa_u$ against the Gaussian functions $e^{-a\tau^2}$,
	$a>0$. The substitution $z=\tau e_\theta+vn_\theta$ in
	\eqref{eq:cumulante-proyectada} gives, under \eqref{eq:integrabilidad},
	\begin{equation}\label{eq:L-A}
		\int_{\R}\kappa_u(\tau)\,e^{-a\tau^2}\,d\tau
		=\sqrt\pi\,\sigma\,L\big(A(\theta,\sigma,a)\big),
	\end{equation}
	where
		$L(A)=\int_{\R^2}\kappa(z)\,e^{-z^\top Az}\,dz,
	\qquad
	A(\theta,\sigma,a)=a\,e_\theta e_\theta^\top
	+\frac{1}{4\sigma^2}\,n_\theta n_\theta^\top.
	$ 
	
	By \eqref{eq:integrabilidad}, $L$ converges
	absolutely on the cone $S_2^{++}$ of $2\times2$ symmetric positive definite
	matrices.

	Intuitively, when $\kappa$ takes both signs, its positive and negative parts
	may cancel in \eqref{eq:cumulante-proyectada} for some projections, and
	Remark~\ref{rem:contraejemplo} below shows that this can happen for any
	finite set of projections fixed in advance. Such cancellations are
	exceptional: by \eqref{eq:L-A}, a failing projection forces the real
	analytic function $L$, which is not identically zero unless $\kappa$ is, to
	vanish, and this can only happen on a null set of projections. A projection
	drawn from an absolutely continuous distribution therefore avoids it with
	probability one, as in the random-projection Cram\'er--Wold theorem
	\citep{CuestaFraimanRansford2007}.
	\begin{proposition}[Sign changes: almost every projection suffices]\label{prop:casi-toda}
		Assume \eqref{eq:integrabilidad} and that $\kappa$ is not almost everywhere
		zero. Then the set $E=\{u\in\Ucal:\ \kappa_u=0 \text{ almost everywhere}\}$
		has Lebesgue measure zero. In particular, so does the set of $u\in\Ucal$
		for which $\Pi_u(S)$ is a Poisson process, which contains
		$\{u:Q_u=Q_u^*\}$.
	\end{proposition}
	
	\begin{proof}
		Since $\kappa_u$ does not depend on $p$,
		$E=\{(\theta,p,\sigma):(\theta,\sigma)\in E'\}$ for some
		$E'\subset(0,\pi)\times\Sigma$. Suppose that $\ell(E)>0$, so that $E'$ has
		positive two-dimensional measure. By \eqref{eq:L-A}, $L$ vanishes at
		$A(\theta,\sigma,a)$ for every $(\theta,\sigma)\in E'$ and $a>0$. The matrix
		$A(\theta,\sigma,a)$ has eigenvectors $e_\theta$ and $n_\theta$, with
		eigenvalues $a$ and $1/4\sigma^2$, so $(\theta,\sigma,a)\mapsto
		A(\theta,\sigma,a)$ is a local diffeomorphism outside the null set
		$\{a=1/4\sigma^2\}$, and $L$ vanishes on a subset of $S_2^{++}$ of positive
		measure.
		
		Now $L(A)=\int e^{-\operatorname{tr}(AM)}\,\nu(dM)$ is the Laplace transform
		of the signed measure $\nu=T_\#(\kappa(z)\,dz)$, with $T(z)=zz^\top$. It is
		therefore real analytic on the open convex cone $S_2^{++}$
		\citep[Ch.~7]{BarndorffNielsen1978}, and, vanishing on a set of positive
		measure, it vanishes identically \citep{MityaginZeroSet2020}. By uniqueness
		of the Laplace transform \citep[Ch.~7]{BarndorffNielsen1978}, $\nu=0$, that
		is, $\int_C\kappa=0$ for every Borel set $C=-C$. If $C$ lies in a
		half-plane through the origin, evenness gives
		$2\int_C\kappa=\int_{C\cup(-C)}\kappa=0$. Hence $\kappa=0$ almost
		everywhere, a contradiction. The last statement follows from
		Lemma~\ref{lem:cumulante-proyectada}.
	\end{proof}
	
	Proposition~\ref{prop:casi-toda} only states that $E$ is negligible. $E$ may
	be nonempty, and a projection fixed in advance may fall into it, as the
	following remark shows.
	
	\begin{remark}[A fixed projection does not suffice when $c_2$ changes sign]
		\label{rem:contraejemplo}
		Let $u_1,\dots,u_k\in\Ucal$ be arbitrary and look for
		$\kappa(z)=P(z)\,e^{-|z|^2}$, with $P$ an even polynomial of degree at most
		$2m$. On the line $\{\tau e_{\theta_j}+vn_{\theta_j}\}$ we have
		$|z|^2=\tau^2+v^2$, so \eqref{eq:cumulante-proyectada} gives
		$\kappa_{u_j}(\tau)=e^{-\tau^2}R_j(\tau)$, where $R_j$ is an even polynomial
		of degree at most $2m$ that depends linearly on $P$. The conditions
		$R_1=\dots=R_k=0$ are $k(m+1)$ linear equations for the $(m+1)^2$
		coefficients of $P$, so for $m=k$ there is a nonzero solution. The
		resulting $\kappa$ is even, smooth and satisfies \eqref{eq:integrabilidad},
		yet $u_1,\dots,u_k\in E$. An explicit example is
		$\kappa(z)=(z_1^2-z_2^2)\,e^{-|z|^2}$: it changes sign under the
		reflections $(z_1,z_2)\mapsto(z_2,z_1)$ and $(z_1,z_2)\mapsto(-z_2,-z_1)$,
		so the integrand in \eqref{eq:cumulante-proyectada} is odd in $v$ for
		$\theta=\pi/4$ and $\theta=3\pi/4$, whatever $\sigma$ and $p$.
		
		By Proposition~\ref{prop:una-proyeccion} and Lemma~\ref{lem:signo}, such a
		$\kappa$ changes sign and is incompatible with Palm domination. The example
		shows that, when the sign of $c_2$ is not known, the vanishing of the
		projected cumulant for finitely many prescribed projections does not imply
		$c_2=0$, and the projection has to be drawn at random. This is the
		point-process counterpart of two classical facts: a function is in general
		not determined by its projections in finitely many directions
		\citep{SmithSolmonWagner1977}, and finitely many projections determine a
		probability distribution only under restrictions such as finite support
		\citep{Renyi1952,Heppes1956}, whereas a single randomly chosen projection
		determines it under suitable moment conditions
		\citep{CuestaFraimanRansford2007}.
	\end{remark}
	
	\begin{remark}[Choice of the kernel]\label{rem:hard-cutoff}
		The two regimes use different properties of the Gaussian kernel. Under a
		fixed sign only its strict positivity matters: every pair of points of $S$
		is seen by the projection. With the hard cylindrical cutoff
		$\1\{|r|\le a\}$, the same argument yields only $\kappa(z)=0$ on the strip
		$|\langle z,n_\theta\rangle|<2a$. When the sign changes, the Gaussian form
		is essential: combined with Gaussian test functions it produces the
		quadratic exponent in \eqref{eq:L-A}, which makes the dependence on
		$(\theta,\sigma)$ real analytic. The discontinuity of the hard cutoff
		destroys this analyticity.
	\end{remark}
	
	\begin{remark}[Higher dimensions]\label{rem:Rd}
		In $\R^d$, $d\ge3$, a projection is determined by a direction
		$e\in S^{d-1}$, a transversal displacement $p\in\R^{d-1}$ and a scale
		$\sigma>0$. Let $N_e$ be a $d\times(d-1)$ matrix whose columns form an
		orthonormal basis of $e^\perp$, chosen measurably in $e$, let
		$P_e=I-ee^\top$, and for $u=(e,p,\sigma)$ set
		\begin{equation}\label{eq:proy-Rd}
			\pi_u(x)=\langle x,e\rangle,
			\qquad
			q_u(x)=w_\sigma\big(|N_e^\top x-p|\big).
		\end{equation}
		The parameter space is $S^{d-1}\times\R^{d-1}\times\Sigma$, with the product
		of surface and Lebesgue measures. Lemma~\ref{lem:cumulante-proyectada} holds
		with $\kappa_u(\tau)=(\sqrt\pi\,\sigma)^{d-1}\int_{\R^{d-1}}
		\kappa(\tau e+N_ev)\,e^{-|v|^2/4\sigma^2}\,dv$, and the proofs of
		Lemma~\ref{lem:signo}, Proposition~\ref{prop:una-proyeccion} and
		Remark~\ref{rem:contraejemplo} carry over verbatim; in the latter, the even
		polynomials of degree at most $2m$ in $d$ variables form a space of
		dimension at least $(m+1)^2$, and the explicit example works with the
		directions $e=(\mathbf e_1\pm\mathbf e_2)/\sqrt2$.
		
		Proposition~\ref{prop:casi-toda} also holds, but its proof needs a
		modification. The matrices $a\,ee^\top+b\,P_e$, with $e\in S^{d-1}$ and
		$a,b>0$, form a family of dimension $d+1<d(d+1)/2$, so they do not cover a
		set of positive measure in $S_d^{++}$. Instead, the function
		$G(e,a,b)=L(a\,ee^\top+b\,P_e)$ is real analytic on the connected manifold
		$S^{d-1}\times(0,\infty)^2$; if $\ell(E)>0$, it vanishes on a set of
		positive measure and hence identically \citep{MityaginZeroSet2020}. For each
		$e$, uniqueness of the Laplace transform in $(a,b)$
		\citep[Ch.~7]{BarndorffNielsen1978} shows that the image of $\kappa(z)\,dz$
		under $z\mapsto(\langle z,e\rangle^2,|P_ez|^2)$ is zero. In polar
		coordinates $z=r\omega$ and by a density argument, this means that, for
		almost every $r$,
		$\int_{S^{d-1}}\kappa(r\omega)\,\phi(\langle\omega,e\rangle^2)\,d\omega=0$
		for every direction $e$ and every bounded Borel function $\phi$ on $[0,1]$.
		Taking $\phi=\1_{[1-\delta,1]}$ and using that $\kappa$ is even,
		$\kappa(r\,\cdot)$ integrates to zero over every spherical cap, and
		Lebesgue's differentiation theorem on the sphere \citep[Ch.~2]{Mattila1995}
		gives $\kappa=0$ almost everywhere. Consequently,
		Theorem~\ref{thm:main-one-projection} holds in $\R^d$.
	\end{remark}

	\section{Identification through a single projection}\label{sec:teorema-principal}
	
	\begin{theorem}[A single projection identifies non-Poissonianity]
		\label{thm:main-one-projection}
		Let $S$ be a simple stationary point process on $\R^2$ with intensity
		$\lambda>0$ and second-order density $\lambda(x,y)$ as in
		\eqref{eq:segundo-orden}, and let $c_2$ and $\kappa$ be as in \eqref{eq:c2},
		with $\kappa$ locally integrable.
		\begin{enumerate}
			\item[(a)] \emph{Palm domination: every projection.} Assume either
			$S\cup\{0\}\preceq S^0$ or $S\cup\{0\}\succeq S^0$. If $S$ is not a stationary Poisson process, then
			\begin{equation}\label{eq:toda-proyeccion}
				Q_u\neq Q_u^*\qquad\text{for every } u\in\Ucal .
			\end{equation}
			\item[(b)] \emph{No sign information: almost every projection.} Assume
			\eqref{eq:integrabilidad} and that $c_2$ is not almost everywhere zero. Then
			\begin{equation}\label{eq:exceptional-zero-main}
				\ell\{u\in\Ucal:\ Q_u=Q_u^*\}=0 .
			\end{equation}
			Consequently, if $U$ is a random element of $\Ucal$ whose distribution is
			absolutely continuous with respect to Lebesgue measure,
			\begin{equation}\label{eq:one-random-projection}
				\PP\{Q_U=Q_U^*\}=0 .
			\end{equation}
		\end{enumerate}
	\end{theorem}
	
	\begin{proof}
		(a) Assume $S\cup\{0\}\preceq S^0$, the other case is proven in the same way. By Lemma~\ref{lem:signo}, $\kappa\ge0$. If $\kappa=0$ almost everywhere,
		then \eqref{6} holds for almost every $(x,y)$, which is all that the
		proof of Proposition~\ref{prop:chen-stein} \new{(Appendix~\ref{app:proofs})} uses, since $\lambda(x,y)$ enters
		it only through integrals; hence $S$ is Poisson, which is
		excluded. Hence $\kappa$ is nonnegative and not almost everywhere zero, and
		Proposition~\ref{prop:una-proyeccion} gives \eqref{eq:toda-proyeccion}.
		
		(b) Equation \eqref{eq:exceptional-zero-main} is
		Proposition~\ref{prop:casi-toda}, and \eqref{eq:one-random-projection}
		follows from the absolute continuity of the law of $U$.
	\end{proof}

	The two parts of the theorem differ in what is assumed and in what is concluded:
	\begin{itemize}
		\item In (a) the projection can be any one, fixed in advance; no
		integrability beyond local integrability is needed; and the conclusion
		covers every non-Poisson process satisfying Palm domination.
		\item In (b) the projection must be random, since by
		Remark~\ref{rem:contraejemplo} every prescribed finite family of projections
		can fail; and the conclusion only covers processes with $c_2\neq0$.
		Non-Poisson processes with $c_2=0$ are outside the reach of second-order
		methods. Under Palm domination they do not exist, by
		Proposition~\ref{prop:chen-stein}.
	\end{itemize}

	\begin{definition}[Second-order Poisson identifiability]\label{def:soi}
		A class $\mathcal C$ of simple stationary point processes is said to be
		\emph{second-order Poisson identifiable} (SOI) if, for every $S\in\mathcal C$,
		$
		c_2=0\quad\text{almost everywhere}
		\qquad\Longrightarrow\qquad
		S\text{ is a stationary Poisson process}$.
		
		Thus, within a SOI class, every non-Poisson process necessarily has
		$c_2\neq0$. Notice that this condition does not impose any sign restriction
		on $c_2$.
	\end{definition}
	
	\begin{corollary}[Random projection identification in SOI classes]
		\label{cor:soi-random-projection}
		Let $S$ satisfy the assumptions of Theorem~\ref{thm:main-one-projection},
		assume \eqref{eq:integrabilidad}, and suppose that $S$ belongs to an SOI
		class. If $S$ is not Poisson, then
		$
		\ell\{u\in\Ucal:Q_u=Q_u^*\}=0.
		$
		Consequently, if $U$ has a distribution absolutely continuous with respect
		to Lebesgue measure on $\Ucal$,
		$
		\PP\{Q_U=Q_U^*\}=0.
		$
	\end{corollary}
	
	\begin{proof}
		Since the class is SOI, non-Poissonianity implies $c_2\neq0$. The conclusion
		is therefore exactly Theorem~\ref{thm:main-one-projection}(b).
	\end{proof}
	
	\begin{remark}[A SOI class beyond Palm domination]\label{rem:soi-cox}
		Stationary Cox processes with finite second moments form a SOI class. Indeed,
		if $S$ is driven by a stationary random intensity field $\Lambda$ with
		$\EE\Lambda(x)=\lambda$, then
		$
		c_2(x,y)=\Cov\{\Lambda(x),\Lambda(y)\}.
		$
		Hence, if $c_2=0$ almost everywhere, then for every bounded Borel set $A$,
		\[
		\Var\!\left(\int_A\Lambda(x)\,dx\right)
		=\int_A\!\int_A c_2(x,y)\,dx\,dy=0.
		\]
		Thus the random directing measure equals $\lambda\,dx$ almost surely (first
		on a countable generating class and then, by uniqueness of measures, on all
		Borel sets), and the Cox process reduces to a stationary Poisson process.
		
		The Cox class is not contained in either Palm-domination regime. For example, let
		$
		\Lambda(x)=\lambda\{1+\varepsilon
		\cos(\omega^\top x+\Phi)\},
		\qquad 0<\varepsilon<1,\quad \omega\neq0
		$,
		where $\Phi\sim\mathrm{Unif}(0,2\pi)$. The random phase makes the field
		stationary, while
		$
		c_2(x,y)=\frac{\lambda^2\varepsilon^2}{2}
		\cos\{\omega^\top(y-x)\}
	$.
		Therefore $c_2$ takes both positive and negative values. By
		Lemma~\ref{lem:signo}, neither Palm domination nor reverse Palm domination
		can hold. Nevertheless, the process belongs to the SOI Cox class, and
		Corollary~\ref{cor:soi-random-projection} shows that almost every Gaussian
		projection identifies its non-Poissonianity.
	\end{remark}

	\begin{remark}[Fixed sign without Palm domination]\label{rem:signo-fijo}
		Proposition~\ref{prop:una-proyeccion} only uses the sign of $c_2$. Hence any
		stationary process with $c_2\ge0$ or $c_2\le0$, not almost everywhere zero,
		is detected by every projection, even without Palm domination. Palm domination is needed only to
		guarantee that a non-Poisson process has $c_2\neq0$. The fixed-sign case covers determinantal processes, which moreover satisfy the reverse Palm domination, since their reduced Palm version is obtained by removing points from the process
		itself \citep{Goldman2010,MollerOReilly2021}, with $\kappa=-C^2\le0$, where $C(x-y)$ denotes the (real, symmetric) kernel of the determinantal process.
	\end{remark}
	
	\begin{remark}[Population versus sample]\label{rem:population-vs-sample}
		Theorem~\ref{thm:main-one-projection} is a statement about distributions.
		It does not assert that a test based on one realization and one projection
		has power one. In finite samples the power depends on $u$; for instance, an
		anisotropic $c_2$ may produce a weak signal $\kappa_u$ in some directions.
		Aggregating several random projections stabilizes the signal and covers
		both parts of the theorem at once. This is the construction of
		Section~\ref{sec:implementacion}.
	\end{remark}
	
	\begin{corollary}\label{cor:fkg-one-projection}
		Let $S$ be as in Theorem~\ref{thm:main-one-projection}. If $S$ is not Poisson and is either {positive or negative} associated, then, by
		Lemma~\ref{lem:fkg-palm} and part (a), every smoothed Gaussian projection
		identifies non-Poissonianity.
	\end{corollary}
	
	\begin{remark}[Scope of the hypotheses]\label{rem:alcance-hipotesis}
		Condition \eqref{eq:integrabilidad} holds whenever $|\kappa|$ is bounded by
		a polynomial. Palm domination is not automatic, but by
		Lemma~\ref{lem:fkg-palm} it holds for stationary positive associated
		processes, and the reverse domination holds for negative associated ones. In a Cox process driven by a random intensity field
		$\Lambda(x)$,
		$
		c_2(x,y)=\operatorname{Cov}\big(\Lambda(x),\Lambda(y)\big)$.
		
		Cox models with nonnegative covariance are therefore natural examples of
		part (a). Under standard conditions these include log-Gaussian Cox
		processes, shot-noise Cox processes and various Poisson cluster processes
		\citep{MollerWaagepetersen2003,DiggleMoragaRowlingsonTaylor2013}.
		More generally, stationary Cox processes with finite second moments are SOI; their covariance may change sign, in which case Palm domination fails but Corollary~\ref{cor:soi-random-projection} still yields identification by almost every random projection; see Remark~\ref{rem:soi-cox}.
	\end{remark}
	
	\begin{remark}
		[Higher dimensions] 
		By Remark~\ref{rem:Rd}, Theorem~\ref{thm:main-one-projection} also holds in
		$\R^d$, $d\ge3$, with the parameter space described there.
	\end{remark}
	\section{The test}\label{sec:implementacion}
	
	We observe $S_A = S\cap A=\{X_1,\dots,X_n\}$ in a bounded window
	$A\subset\mathbb R^2$ and test
	$H_0$: $S$ is a homogeneous Poisson process on $\mathbb R^2$,
	estimating the intensity by $\hat\lambda = n/|A|$.
	
	\paragraph{One projection.}
	Fix $u=(\theta,p,\sigma)$. Retain each $X_i$ independently with probability
	$q_u(X_i)$, and let $Y_1,\dots,Y_{N_u}$ be the projected coordinates
	$\pi_u(X_i)$ of the retained points. Because of the window, the projected
	process is not homogeneous even under $H_0$; its estimated mean measure is
	$
	\hat\Lambda_{A,u}(t)=\hat\lambda\int_A
	\mathbf 1\{\langle x,e_\theta\rangle\le t\}\,
	w_\sigma\big(\langle x,n_\theta\rangle-p\big)\,dx$, 
	which we approximate numerically on a regular grid over $A$. Under $H_0$
	the retained projected points form a Poisson process on $\R$ with
	continuous mean measure $\Lambda_{A,u}$. By the Mapping Theorem
	\citep[\S2.3]{kingman1993}, up to the estimation of $\lambda$, the points
	$Z_j=\hat\Lambda_{A,u}(Y_j)$ form a homogeneous rate-one Poisson process on
	$[0,L_u]$, with $L_u=\hat\Lambda_{A,u}(\infty)$.
	
	Any one-dimensional discrepancy $D_u$ may then be applied to
	$\{Z_j\}$: spacings compared with the unit exponential \citep{Pyke1965},
	counts in subintervals, Laplace functionals, or void probabilities. We use
	counts: split $[0,L_u]$ into $K_u$ cells with expected count
	$E_u=L_u/K_u$ and observed counts $O_1,\dots,O_{K_u}$, and set
	\[
	D_u = 2\sum_{k=1}^{K_u}\Big(O_k\log\frac{O_k}{E_u}-(O_k-E_u)\Big),
	\qquad 0\log 0 = 0 ,
	\]
	the Poisson deviance. When the total count is held fixed it reduces to the
	likelihood-ratio statistic $G^2$, the member of index $0$ of the
	power-divergence family \citep{CressieRead1984}.
	Under the Palm domination $S\cup\{0\}\preceq S^0$, $c_2\ge 0$ (Lemma~\ref{lem:signo}), so departures
	from $H_0$ show up as
	overdispersion of the projected counts, and only large values of $D_u$ are
	evidence against $H_0$. \new{For repulsive alternatives, $c_2\le0$, the departure appears instead as underdispersion, which coarse cell counts barely register; replacing the deviance by a Greenwood spacings score and the maximum over projections by the average yields the variant studied in Appendix~\ref{subsec:sim-repulsive}.}
	
	\paragraph{Several projections and calibration.}
	At the population level, under Palm domination any single projection
	identifies non-Poissonianity (Theorem~\ref{thm:main-one-projection}(a)).
	Without sign information, the projection must be random
	(Theorem~\ref{thm:main-one-projection}(b)). We therefore draw $B$
	projections $u_1,\dots,u_B$ i.i.d.\ from an absolutely continuous
	distribution on $\Ucal$. This covers both cases and, in finite samples,
	stabilizes the signal. The projections are aggregated through
	$
	T\new{{}=T_{\mathrm{GT}}}=\max_{1\le b\le B} D_{u_b}
	$\new{, which we call the Gauss--thin statistic}.
	The null distribution of $T$ is obtained by parametric simulation: generate
	$M$ homogeneous Poisson patterns in $A$ with intensity $\hat\lambda$,
	recompute $T^{(1)},\dots,T^{(M)}$ with the \emph{same} projections
	(re-estimating $\lambda$ in each replicate), and report
	\[
	\hat p_{\mathrm{val}}
	=\frac{1+\#\{m\le M:\ T^{(m)}\ge T_{\mathrm{obs}}\}}{M+1}.
	\]
	This is a Monte Carlo test in the sense of \citet{BesagDiggle1977}; see also
	\citet{BaddeleyRubakTurner2015}. The calibration is conditional on the chosen
	projections and absorbs, at once, the estimation of $\lambda$, the
	aggregation over projections and the randomness of the thinning.
	
	In \textsf{R}, the window, the grid approximation of $\hat\Lambda_{A,u}$ and
	the simulation under $H_0$ can be handled with \texttt{spatstat}
	\citep{spatstat}, and other one-dimensional discrepancies, such as the
	Cram\'er--von Mises and Anderson--Darling statistics, are available in
	\texttt{goftest} \citep{goftest}; see also
	\citet{Diggle2013,MollerWaagepetersen2003}.
	
	\begin{algorithm}[ht]
		\caption{Poissonianity test based on smoothed Gaussian projections}
		\label{alg:poisson-projections}
		\begin{algorithmic}[1]
			\Statex \textbf{Input:} pattern $S_A\subset A$, number of projections $B$,
			null simulations $M$, scale range $\Sigma$.
			\Statex \textbf{Output:} $T_{\mathrm{obs}}$ and $\hat p_{\mathrm{val}}$.
			\State Draw $u_b=(\theta_b,p_b,\sigma_b)$, $b=1,\dots,B$, with
			$\sigma_b\in\Sigma$, i.i.d.\ from the distribution on $\Ucal$
			described above.
			\State For a pattern $X\subset A$, define $T(X)=\max_b D_{u_b}(X)$, where
			$D_u(X)$ is obtained by thinning with $q_u$, projecting with
			$\pi_u$, transforming by $\hat\Lambda_{A,u}$ and counting.
			\State $T_{\mathrm{obs}}\gets T(S_A)$.
			\For{$m=1,\dots,M$}
			\State Simulate a homogeneous Poisson pattern $X^{(m)}$ in $A$ with
			intensity $\hat\lambda$.
			\State $T^{(m)}\gets T(X^{(m)})$.
			\EndFor
			\State \Return $T_{\mathrm{obs}}$ and $\hat p_{\mathrm{val}}$.
		\end{algorithmic}
	\end{algorithm}
	
	In $\R^d$ the procedure is the same, with the notation of
	Remark~\ref{rem:Rd}: $u=(e,p,\sigma)$ with $e\in S^{d-1}$ and
	$p\in\R^{d-1}$, and $q_u$ as in \eqref{eq:proy-Rd}.

	\section{Real Data: impact craters and the dating of planetary surfaces}
	\label{sec:craters}
	
\begin{figure}[!ht]
		\centering
		\includegraphics[width=1\textwidth]{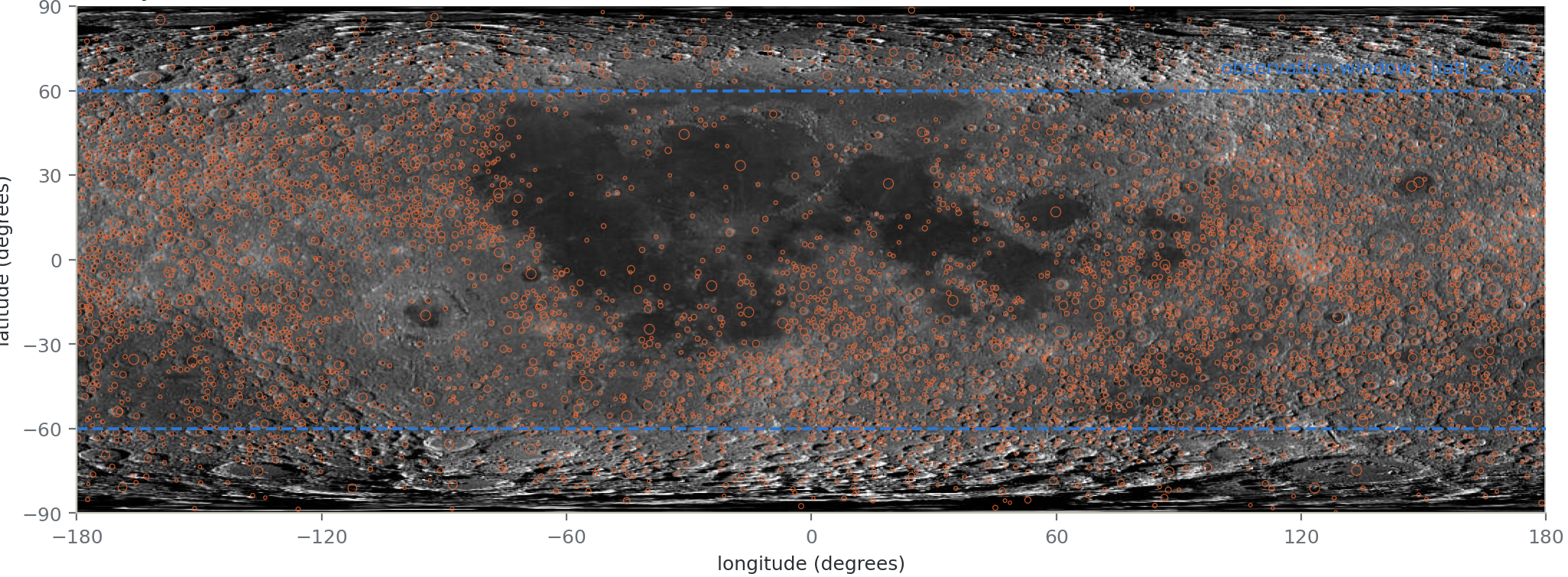}
		\caption{\new{Lunar ray systems and secondary crater swarms.} {Catalogued craters (circles) over a global lunar basemap; the dashed lines bound the band $|\varphi|\le 60^\circ$ used in the analysis. Image: NASA/GSFC/Arizona State University.}}
		\label{fig:moon}
	\end{figure}

\begin{figure}[htbp]
	\centering
	\includegraphics[width=1\textwidth]{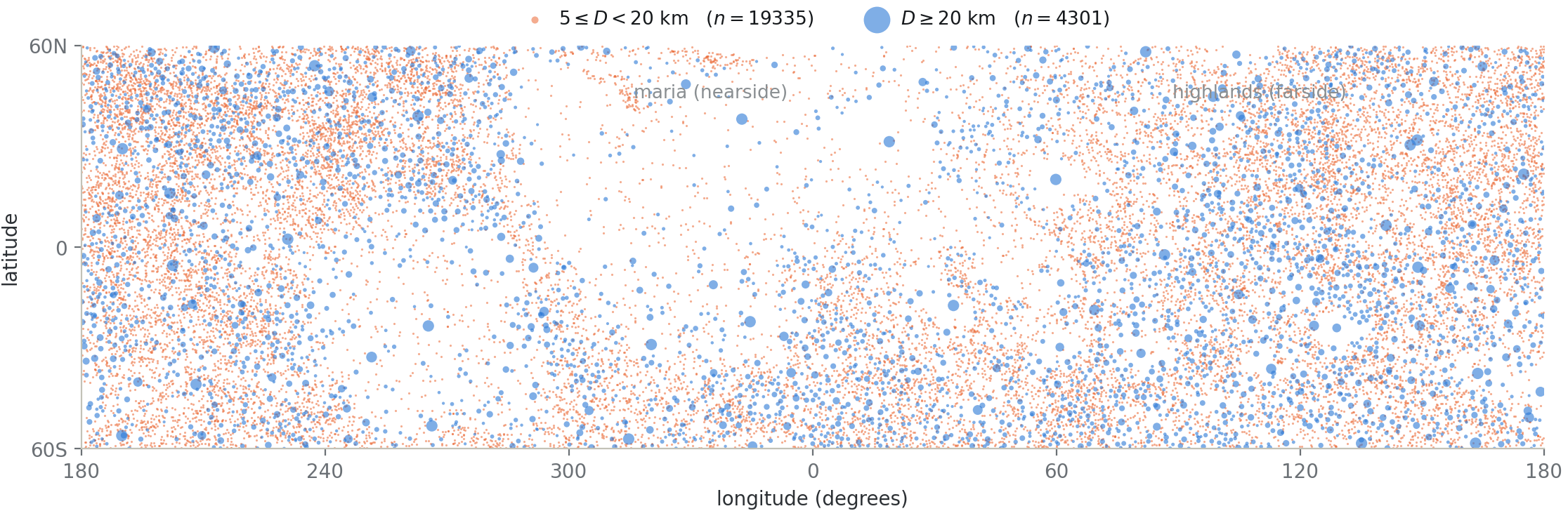}
	\caption {Spatial distribution of lunar craters.}
	\label{fig:points}
\end{figure}

	Crater chronology estimates planetary surface ages without physical samples {\citep{neukum2001}}: under a constant impact flux, crater density is proportional to surface exposure time. Primary impacts form a homogeneous Poisson process, and the Colouring Theorem {\citep{kingman1993}} guarantees that filtering by diameter preserves this property, since diameter is an independent mark. However, large impacts eject debris that forms clustered \emph{secondary craters} {\citep{mcewen2006}} (e.g., ray systems visible in \new{Fig.~\ref{fig:moon}}), which bias age estimates upward and inflate count variance. The lunar crust comprises two distinct terrains: the dark, basaltic \emph{maria} ($\approx 3.5$~Ga, mainly nearside) and the bright, heavily cratered \emph{highlands} ($\approx 4.4$~Ga, mainly farside). Because crater density scales with age, these terrains differ in density by nearly an order of magnitude, creating a sharp boundary rather than a gradient (Fig.~\ref{fig:points}). {\citet{michael2012}} state that reliable surface dating requires excluding diameter ranges that exhibit spatial non-randomness. They work with prespecified diameter intervals and nearest-neighbour statistics.
To apply our planar test on $\mathbb{R}^2$ to spherical data, we map the latitude band {$\mathcal B=\{|\varphi|\le\varphi_{\max}\}$, where $\varphi$ denotes latitude,} onto a rectangle $A$ via a rescaled Lambert cylindrical equal-area projection {\citep{lambert1772,snyder1987}}. Because this projection is area-preserving, the Mapping Theorem transforms a Poisson process on {$\mathcal B$} into one on $A$, ensuring an exact test level without edge corrections. {Our procedure relies on linear projections, so metric distortion may reduce its power but does not alter its level.} Combining two lunar catalogues {\citep{head2010,povilaitis2018}} within the common band {$\varphi_{\max}=60^\circ$} yields {$n=23{,}636$} craters with $D \ge 5$~km.
Testing for homogeneity here is invalid because observed crater density reflects both impact flux {$F$} and local surface age {$\mathrm{age}(x)$};	{$\lambda(x)=F\cdot\mathrm{age}(x)$}.	Since {$\mathrm{age}(x)$} varies sharply between maria and highlands ({Fig.~\ref{fig:points}}), the proper null model is an \emph{inhomogeneous} Poisson process with intensity $\lambda(x)$. Rejecting this null isolates true spatial clustering rather than geological age differences. Because the Colouring and Mapping Theorems hold for inhomogeneous processes, Algorithm~1 adapts directly: Thinning with retention probability $q_u(x)$ produces a Poisson process of intensity $\lambda(x)q_u(x)$;	Projecting onto $\mathbb{R}$ gives a Poisson process with cumulative intensity
\begin{equation}
	\Lambda_{A,u}(t)=\int_A \mathbf{1}\{\pi_u(x)\le t\}\,q_u(x)\,\lambda(x)\,dx;
	\label{eq:lambda-inhom}
\end{equation}
transforming via $\Lambda_{A,u}$ yields a unit-rate Poisson process on {$[0,L_u]$}, where {$L_u=\Lambda_{A,u}(\infty)$}.
Thus, $\widehat\lambda$ simply enters inside \eqref{eq:lambda-inhom} instead of factoring out as $n/|A|$, and is re-estimated in each Monte Carlo replicate to account for estimation error. Note that this tests Poissonianity conditional on $\widehat\lambda$, without extending the stationarity-based identification of {Theorem~\ref{thm:main-one-projection}}. Estimating $\lambda$ from the data introduces Bartlett's identifiability issue {\citep{bartlett1964}}, making the results dependent on the smoothing bandwidth $h$. We evaluate five {approximately} doubling bandwidths, $h\in\{90,181,361,752,1505\}$~km, bounded by physical scales: below $\approx 90$~km (three times the smallest mean nearest-neighbour distance), $\widehat\lambda$ overfits individual craters, so a non-rejection there is not evidence of Poissonianity; above $\approx 1505$~km ($41\%$ of the window width), $\widehat\lambda$ becomes nearly constant, recovering the homogeneous test. The intermediate range forms the \emph{geological window}, capturing macro-terrain features without absorbing local clustering. We set $B=20$ projections, $\sigma_b \in \Sigma=\{0.05,\,0.08,\,0.12,\,0.18,\,0.25\}$ (uniformly drawn relative to the short side of $A$), $M=499$ simulations, and $\approx 8$ expected points per cell. On simulated inhomogeneous data with a $19{:}1$ intensity contrast, the empirical rejection rate is $0.042$ (nominal $0.05$), rising to a power of $0.983$ under added Cox clustering.

\begin{table}[t]
	\centering
	\caption{$p$-values of $T_{\mathrm{GT}}$ under the inhomogeneous Poisson null across bandwidths $h$ and diameter windows. Rejections at the $5\%$ level are in boldface.}
	\label{tab:craters}
	\begin{tabular}{lccccc}
		\toprule
		& \multicolumn{5}{c}{diameter window (km)} \\
		\cmidrule(lr){2-6}
		$h$ (km) & $5$--$7$ & $10$--$14$ & $20$--$30$ & $45$--$70$ & $\ge 120$ \\
		& \small $n=6952$ & \small $n=4125$ & \small $n=1579$ & \small $n=849$ & \small $n=167$ \\
		\midrule
		$90$    & $0.312$          & $0.214$          & $0.890$          & $0.752$ & $0.446$ \\
		\addlinespace[2pt]
		$181$   & $\mathbf{0.002}$ & $0.676$          & $0.710$          & $0.496$ & $0.448$ \\
		$361$   & $\mathbf{0.002}$ & $\mathbf{0.010}$ & $\mathbf{0.012}$ & $0.112$ & $0.508$ \\
		$752$   & $\mathbf{0.002}$ & $\mathbf{0.002}$ & $\mathbf{0.002}$ & $0.432$ & $0.366$ \\
		\addlinespace[2pt]
		$1505$  & $\mathbf{0.002}$ & $\mathbf{0.002}$ & $\mathbf{0.002}$ & $0.076$ & $\mathbf{0.002}$ \\
		\bottomrule
	\end{tabular}
\end{table}
In Table~\ref{tab:craters}, $h=1505$~km rejects almost everywhere due to the maria–highlands contrast, whereas $h=90$~km never rejects due to overfitting. Within the geological window ($181\le h\le 752$~km), craters with $D\ge 45$~km conform to the inhomogeneous Poisson null ($p \ge 0.112$), while those with $D<30$~km reject it for $h\ge 361$~km ($p\le 0.012$); at $h=181$~km only the $5$--$7$~km window does. This places the transition between $30$ and $45$~km, consistent with the spatial scale of secondary crater fields or mutual obliteration.
In practice, the dispersion index $I$, the variance-to-mean ratio of crater counts, drops from $I=13.3$ (for $5$--$7$~km) to $I=0.96$ (for $D\ge 120$~km). Consequently, age uncertainties derived from $5$--$7$~km craters are underestimated by a factor of $\sqrt{I}\approx 3.6$, whereas for $D\ge 45$~km the counts are consistent with independence and no such correction is needed. The diameter threshold called for by \citet{michael2012} is therefore estimated from the data rather than fixed in advance.

\section{Competing interests}
No competing interest is declared.

\section{Author contributions statement}

P.A.F., R.F., S.G., and L.M. contributed equally to all aspects of this work and approved the final manuscript. Authors are listed in alphabetical order.

\section{Acknowledgments}
We thank Jorge Busch and Pablo Groisman for discussions at an early stage of this research.

\clearpage
\appendix
\numberwithin{equation}{section}
\numberwithin{figure}{section}
\numberwithin{table}{section}
\setcounter{theorem}{0}
\renewcommand{\thetheorem}{\thesection.\arabic{theorem}}
\setcounter{remark}{0}
\renewcommand{\theremark}{\thesection.\arabic{remark}}
\setcounter{definition}{0}
\renewcommand{\thedefinition}{\thesection.\arabic{definition}}
\setcounter{example}{0}
\renewcommand{\theexample}{\thesection.\arabic{example}}
\section{Some proofs}\label{app:proofs}
\new{Throughout this section we use the notation of Section~\ref{sec:identificacion} of this article: $\Lambda\subset\R^d$ is a compact set, $S^x$ is the Palm version of $S$ at $x$, $W=|S\cap\Lambda|$ and $W_x=|S^x\setminus\{x\}\cap\Lambda|$.}
\begin{proof}[Proof of Proposition~\ref{prop:chen-stein}]              
		\new{Proposition~\ref{prop:chen-stein} states that a stationary process satisfying \eqref{14} and \eqref{6} is a Poisson process.} We follow the proof of Theorem 24.3 in \citet{lindvall1992}, and use the Palm domination to show that $S$ is Poisson. Assume $S\cup\{0\}\preceq S^0$. Since the process is stationary, this implies that for each $x$ there is a coupling $(S,S^x)$ with marginals distributed as $S$ and $S^x$,  satisfying
		\begin{align}
			\label{17}
			S\cup\{x\}\subset S^x.
		\end{align}
		Denote $\PP$, $\EE$ the distribution of the coupling. 	
		Let $f$ be the solution of the Stein equation for the void probability,
		\begin{align}
			\1\{k=0\}-e^{-\lambda\Leb(\Lambda)}
			=\lambda\Leb(\Lambda)f(k+1)-kf(k),
			\qquad k\in\Z_+, \label{3}
		\end{align}
		with $f(0)=0$. Denote
		\begin{align}
			\label{1}
			\|\Delta f\|_\infty:= \sup_{k\ge0}|f(k+1)-f(k)| \leq 1\wedge [\lambda\Leb(\Lambda)]^{-1}.
		\end{align}
		Taking expectations and absolute values in \eqref{3}, 
		\begin{align}
			&\bigl|\PP(W=0)-e^{-\lambda\Leb(\Lambda)}\bigr|\\
			&\quad=
			\bigl|{\EE\bigl(\lambda\Leb(\Lambda)f(W+1)\bigr)-\EE\bigl(Wf(W)\bigr)}\bigr|\\
			&\quad\leq
			\lambda\int_\Lambda
			\EE\left|f(W+1)-f(W_x+1)\right|dx \label{4}\\
			&\quad\leq
			\lambda\|\Delta f\|_\infty
			\int_\Lambda \EE|W-W_x|\,dx \label{2}\\
			&\quad=
			\lambda\|\Delta f\|_\infty
			\int_\Lambda \EE(W_x-W)\,dx
			=0.\label{5}
		\end{align}
		where \eqref{4} follows from \new{\eqref{9} (Lemma~\ref{13})},
		the first identity in \eqref{5} follows from \eqref{17}, and the second one from \new{\eqref{20} (Lemma~\ref{lem:varianza})}.
		We conclude that
		\[
		\PP(|S\cap\Lambda|=0)=e^{-\lambda\Leb(\Lambda)}
		\]
		for every compact $\Lambda$. Void probabilities characterize the distribution of the Poisson process; hence $S$ is a stationary Poisson process of intensity $\lambda$.
		The proof under the 
		reverse domination $S^0\preceq S\cup\{0\}$ uses a coupling satisfying
		$S^x\subset S\cup\{x\}$, which gives $W_x\le W$, so that
		$\EE|W-W_x|=\EE W-\EE W_x=0$ in \eqref{5}. This is the
		negatively related case of the Stein--Chen coupling method \citep{BarbourHolstJanson1992}.
\end{proof}

\begin{proof}[Proof of Lemma~\ref{lem:fkg-palm}]
		\new{Lemma~\ref{lem:fkg-palm} states that positive association implies \eqref{22} and negative association implies \eqref{23}. We start proving  the first implication, \eqref{22}.} It suffices to verify the \new{stochastic order} for functions $f$ that are local, increasing, bounded, and continuous with respect to the vague topology, since this class determines the stochastic order on the space of locally finite configurations; the general case follows by monotone approximation.		
		Let $B_n$ be the closed ball with center $0$ and \new{radius} $1/n$, and define
		$g_n(S):= 1\{|S\cap B_n|\ge 1\}$.
		The functions $g_n$ and $f$ are local and increasing. Positive association implies
		\[
		\EE f(S) \le \frac{\EE(f(S)\,g_n(S))}{\EE g_n(S)}\;\underset{n\to \infty}{\longrightarrow}\; 	\EE f(S^0),
		\]
		where the limit is Eq.\,(9.10) in \citet{last-penrose}. This shows \new{\eqref{22}}.

		To show the second one, \new{\eqref{23}}
		observe that $f(\cdot\setminus B_n)$ and $g_n$ have disjoint supports, and write
		\begin{align}
			\EE f(S\setminus B_n) \ge \frac{\EE(f(S\setminus B_n)\,g_n(S))}{\EE g_n(S)}
			\;\underset{n\to \infty}{\longrightarrow}\; 	\EE f(S^0\setminus\{0\}). 
		\end{align}
		The limit is obtained as in the previous case, using in addition that, since $S$ is simple with finite intensity, $\PP(|S\cap B_n|\ge 2)=o\big(\PP(|S\cap B_n|\ge 1)\big)$: conditionally on $g_n(S)=1$, with probability tending to one $S\setminus B_n$ is $S$ without its unique point in $B_n$, and the continuity of $f$ controls the vanishing shift that brings that point to the origin.
		To conclude, observe that $\EE f(S\setminus B_n) \underset{n\to \infty}{\longrightarrow}\; 	\EE f(S\setminus\{0\}) =\EE f(S)$, and that $S \succeq S^0\setminus \{0\}$ is equivalent to $S\cup\{0\} \succeq S^0$, by the coupling alternative definition of stochastic domination.
\end{proof}	
	\begin{proof}[Proof of Lemma~\ref{lem:cumulante-proyectada}]
		\new{Lemma~\ref{lem:cumulante-proyectada} states that, if $\kappa$ has constant sign or satisfies \eqref{eq:integrabilidad}, the second factorial cumulant density of $\Pi_u(S)$ is given by \eqref{eq:cumulante-proyectada}.} Let $N_u(I)=|\Pi_u(S)\cap I|$. Conditionally on $S$, the retention variables
		are independent, so by \new{\eqref{eq:segundo-orden}} and the elementary properties of independent thinnings \citep[Ch.~5]{last-penrose}, for
		disjoint bounded intervals $I,J$,
		\[
		\Cov\big(N_u(I),N_u(J)\big)
		=\int\!\!\int \1_I(\pi_u x)\,\1_J(\pi_u y)\,
		q_u(x)q_u(y)\,\kappa(y-x)\,dx\,dy .
		\]
		Write $x=te_\theta+rn_\theta$ and $y=se_\theta+r'n_\theta$, so that
		$q_u(x)=w_\sigma(r-p)$, $q_u(y)=w_\sigma(r'-p)$ and
		$y-x=(s-t)e_\theta+(r'-r)n_\theta$. Integrating in $r$ with $v=r'-r$ fixed,
		\[
		\int_{\R} w_\sigma(r-p)\,w_\sigma(r+v-p)\,dr
		=\sqrt{\pi}\,\sigma\,e^{-v^2/4\sigma^2},
		\]
		which gives \new{\eqref{eq:cumulante-proyectada}}. 
		Fubini's theorem applies by
		Tonelli's theorem if $\kappa$ has constant sign, and by \new{\eqref{eq:integrabilidad}} otherwise, because
		$e^{-v^2/4\sigma^2}\le e^{\varepsilon R^2}e^{-\varepsilon(\tau^2+v^2)}$ for
		$|\tau|\le R$ and $\varepsilon\le 1/4\sigma^2$. If $\Pi_u(S)$ is Poisson,
		its counts on disjoint intervals are independent, so the covariance vanishes
		and $\kappa_u=0$ almost everywhere.
\end{proof}

\begin{proof}[Proof of Lemma~\ref{lem:signo}]
	\new{Lemma~\ref{lem:signo} states that $S\cup\{0\}\preceq S^0$ implies $\kappa\ge0$ almost everywhere, and $S^0\preceq S\cup\{0\}$ implies $\kappa\le0$ almost everywhere.} Let $\Lambda$ be a bounded Borel set and $f_k(S)=\min(|S\cap\Lambda|,k)$.
	Each $f_k$ is local and increasing, so the domination $S\cup\{0\}\preceq S^0$ and monotone convergence
	give
	\begin{align}
		\label{24}
		\EE|(S\cup\{0\})\cap\Lambda|\le \EE|S^0\cap\Lambda|.
	\end{align}
	The
	left hand side of \eqref{24} equals $\lambda\,\ell(\Lambda)+\1\{0\in\Lambda\}$. By the
	Campbell--Mecke formula \new{\eqref{7}} applied to
	$f(x,S)=\sum_{y\in S\setminus\{0\}}\1\{x\in A,\ y\in\Lambda\}$, with
	$A$ a bounded set of positive measure, and by \new{\eqref{eq:segundo-orden}} \citep[Ch.~9]{last-penrose}, the right hand side of \eqref{24} is given by
	\[
	\1\{0\in\Lambda\}+\frac1\lambda\int_\Lambda \lambda(0,y)\,dy
	=\1\{0\in\Lambda\}+\lambda\,\ell(\Lambda)
	+\frac1\lambda\int_\Lambda \kappa(y)\,dy .
	\]
	Hence $\int_\Lambda\kappa\ge 0$ for every bounded $\Lambda$, that is,
	$\kappa\ge0$ almost everywhere. The reverse case is identical.
\end{proof}

	\section{Simulation study}
\label{app:simulations}
\label{sec:simulations}
The purpose of this section is to delimit the kinds of departure from the
Poisson hypothesis to which the Gaussian projection procedure is sensitive,
and the kinds to which it is not. The study is organized in two parts. The
first, \S\,\ref{subsec:sim-methods}--\ref{subsec:sim-results}, compares the
proposed statistic with three benchmark procedures on six alternatives
exhibiting aggregation or intensity heterogeneity, all of them with
$c_2\ge0$. The second, \S\,\ref{subsec:sim-repulsive}, treats the repulsive
case $c_2\le0$ separately; it calls for a change of orientation in the
statistics and admits a reading of its own.
All simulations are carried out in the window $\new{A}=[0,1]^2$, under the null
hypothesis that $S$ is a homogeneous Poisson process of intensity
$\lambda_0$ or, equivalently, that $n\sim\Poi(\lambda_0)$ and that
$X_1,\ldots,X_n\mid n$ are independent and uniform on $\new{A}$. Each alternative
family is indexed by a single parameter $\tau\in[0,1]$, with $\tau=0$ giving
the null model and increasing values of $\tau$ intensifying the structure
that violates it.
All procedures are calibrated in the same way. For a method $j$ we generate
$M$ homogeneous Poisson patterns on $\new{A}$ with intensity
$\widehat\lambda=n/|\new{A}|$ and obtain the statistics
$T_j^{(1)},\ldots,T_j^{(M)}$, re-estimating $\lambda$ in each replicate and
reusing the same projections $u_1,\ldots,u_B$ that were applied to the
observed pattern; the critical value is the empirical quantile
$c_{\alpha,j}=\widehat Q_{1-\alpha}\{T_j^{(1)},\ldots,T_j^{(M)}\}$. Every
statistic is oriented so that large values indicate departure from the null,
so that the rejection rule is always one-sided in the upper tail,
$T_{j,\mathrm{obs}}>c_{\alpha,j}$, and the empirical power is estimated over
$R$ replicates of the alternative by
\[
\widehat\beta_j(\tau)=\frac1R\sum_{r=1}^R\1\{T_{j,r}(\tau)>c_{\alpha,j}\}.
\]
Under this scheme the reference distribution absorbs, simultaneously, the
estimation of $\lambda$, the aggregation over projections and the randomness
of the thinning.
The projections $u_b=(\theta_b,p_b,\sigma_b)$, $b=1,\ldots,B$, are generated
with directions $\theta_b$ uniform on $(0,\pi)$, offsets $p_b$ uniform on the
range of the transversal coordinate of $\new{A}$, and bandwidths $\sigma_b$ drawn
with equal probability from $\{0.08,0.12,0.18,0.25\}$. It is worth recalling
the convention of \new{Section~\ref{sec:proyecciones}}: with $e_\theta=(\cos\theta,\sin\theta)$ and
$n_\theta=(-\sin\theta,\cos\theta)$,
\begin{equation}
	\label{eq:sim-qu}
	\pi_u(x)=\langle x,e_\theta\rangle,
	\qquad
	q_u(x)=w_\sigma\!\left(\langle x,n_\theta\rangle-p\right),
	\qquad
	w_\sigma(r)=e^{-r^2/2\sigma^2},
\end{equation}
that is, the thinning depends on the \emph{transversal} distance from the
point to the line $L_{\theta,p}$, and not on its projected coordinate.
The \new{cumulative intensity $\hat\Lambda_{A,u}(t)=\widehat\lambda\int_A\1\{\pi_u(x)\le t\}q_u(x)\,dx$ of Section~\ref{sec:implementacion}} is approximated
once per projection on a regular grid, the transformation of
Algorithm~\ref{alg:poisson-projections} is $Z_i=\new{\hat\Lambda_{A,u}}(Y_i)$, and the
transformed interval is divided into between four and ten cells, keeping
about eight expected points per cell. The results reported below correspond
to $\lambda_0=300$, $B=20$, $M=800$, $R=300$ and nominal level
$\alpha=0.05$; with $R=300$ the Monte Carlo standard error of the powers is
at most $0.029$.
\subsection{The proposed statistic}
\label{subsec:sim-methods}
\label{subsec:sim-our-stat}
\new{The statistic $T_{\GT}$ of Section~\ref{sec:implementacion} (Algorithm~\ref{alg:poisson-projections}) is the Gauss--thin statistic.} For each projection $u_b=(\theta_b,p_b,\sigma_b)$
the Gaussian thinning is applied, the retained points are projected and
transformed by the estimated cumulative intensity, so that under the null
hypothesis the result behaves approximately as a homogeneous Poisson process
of unit rate on $[0,\new{L_{u_b}}]$. Dividing that interval into $K_b$ cells, we
use the Poisson deviance \citep{CressieRead1984},
\[
D_b=2\sum_{k=1}^{K_b}
\left\{O_{bk}\log\!\left(\frac{O_{bk}}{E_{bk}}\right)-(O_{bk}-E_{bk})\right\},
\qquad
E_{bk}=\frac{\new{L_{u_b}}}{K_b},
\]
with the convention $0\log(0/E_{bk})=0$, and the discrepancies are aggregated, \new{with $D_b=D_{u_b}$,} through
\begin{equation}
	\label{eq:sim-main-stat}
	T_\GT=\max_{1\le b\le B} D_b .
\end{equation}
The level of this test does not come from the $\chi^2$ approximation to the
deviance, but from the joint calibration just described.
$T_\GT$ combines three separable choices: the Gaussian geometry, the
univariate discrepancy applied to the transformed points, and the rule that
aggregates the $B$ projections. In \S\,\ref{subsec:sim-repulsive} we show
that regularity forces a variant of the last two.
\subsection{Benchmark tests}
\label{subsec:sim-competitors}
We consider three benchmark statistics, which respond to different detection
mechanisms: structure visible through projections, coarse-scale intensity
heterogeneity, and local proximity between points.
The first, $T_{\mathrm{PIT}}$, is a deliberately elementary construction
based on one-dimensional projections and the probability integral transform,
dispensing with the thinning. It is \emph{ad hoc} rather than a procedure
taken from the literature, although each of its pieces has known antecedents:
the combination of projection and probability integral transform follows the
spirit of goodness-of-fit testing through random projections
\citep{CuestaFraimanRansford2006,CuestaFraimanRansford2007}, the count excess
is of scan type \citep{Naus1965}, and the minimum-spacing statistic comes
from the classical theory of spacings \citep{Pyke1965}. For each direction
$\theta$ we project $Y_i^{(\theta)}=\langle X_i,e_\theta\rangle$ and apply
$P_i^{(\theta)}=F_\theta(Y_i^{(\theta)})$, where $F_\theta$ is the
distribution of $\langle X,e_\theta\rangle$ for $X$ uniform on $\new{A}$, so that
under the null $P_i^{(\theta)}$ is uniform on $[0,1]$. We take the maximum,
over directions and over both components, of
\[
S_h(\theta)=\frac{\max_t\#\{i:P_i^{(\theta)}\in[t,t+h]\}-nh}{\{nh(1-h)\}^{1/2}},
\quad h\in\{0.05,0.10,0.20\},
\]
\[
S_{\mathrm{gap}}(\theta)=-\log\!\big[n^2\min_i\{P^{(\theta)}_{(i+1)}-P^{(\theta)}_{(i)}\}\big].
\]
The role of $T_{\mathrm{PIT}}$ is to quantify how much is lost by dispensing
with the thinning, and the difference from $T_\GT$ is not merely empirical.
Theorem~\ref{thm:main-one-projection} rests on the Gaussian weight $q_u$:
it makes the projected cumulant of Lemma~\ref{lem:cumulante-proyectada}
well defined for the whole homogeneous process and, in part (b), real
analytic on an open set of parameters $(\theta,\sigma)$
(see \new{\eqref{eq:L-A}}); without thinning that open set
collapses to the direction $\theta$ alone, and the genericity argument
loses its support. No
identification result in $\R^2$ is therefore available for the plain
orthogonal projection.
The second, $T_{\mathrm{quad}}$, is the classical quadrat count test
\citep{Diggle2013,BaddeleyRubakTurner2015}: it divides $\new{A}$ into four quadrats
$C_1,\ldots,C_4$ and compares the observed counts with those expected under
homogeneity, $T_{\mathrm{quad}}=\sum_{j=1}^{4}(O_j-n/4)^2/(n/4)$ with
$O_j=\#(S_{\new{A}}\cap C_j)$. It operates at a fixed scale and is sensitive to
intensity differences between large regions.
The third, $T_{\mathrm{CE}}$, is based on the nearest-neighbour distance of
\citet{ClarkEvans1954}. With $R_i=\min_{k\neq i}\|X_i-X_k\|$
and $\overline R=n^{-1}\sum_iR_i$ we use $T_{\mathrm{CE}}=1/\overline R$, so
that large values correspond to stronger local aggregation. For all three
procedures the null distribution is obtained by the same Monte Carlo
calibration, so that the comparison depends neither on the usual $\chi^2$
approximation for $T_{\mathrm{quad}}$ nor on the classical asymptotic
normalization of the Clark--Evans index.
\subsection{Alternatives}
\label{subsec:sim-scenarios}
\begin{figure}[htbp]
	\centering
	\includegraphics[width=\textwidth]{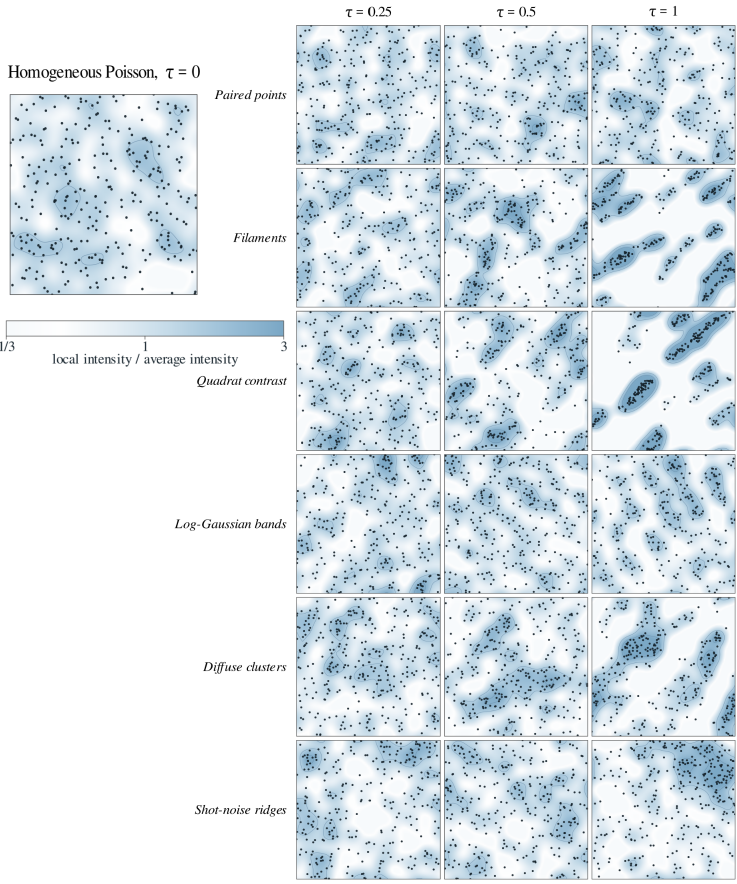}
	\caption{Simulated patterns at $\tau\in\{0,0.25,0.5,1\}$ for
		{the six alternatives of \S\,\ref{subsec:sim-scenarios}; rows, from top:
		Pairs, Filaments, Contrast, LGCP bands, Diffuse clusters and Shot-noise ridges.}}
	\label{fig:patterns-d2}
\end{figure}

The six alternatives of this first part all have $c_2\ge0$. The aim is not to
produce a ranking, but to make the study operate as a series of positive and
negative controls according to the type of departure. The first three
separate different kinds of signal in a controlled way, by fixing which
aspect of the pattern remains identical to that of the null model; the
remaining three are of Cox type and contribute random heterogeneity of the
intensity.
\begin{itemize}
		\item In the {\emph{Pairs}} scenario, $\new{A}$ is divided into quadrats
	and independent counts $N_j\sim\Poi(\lambda_0/4)$ are preserved exactly as
	under the null. Conditionally on $N_j$, a proportion $\tau$ of the points is
	rearranged into close pairs: if $Z\sim\Ucal(C_j)$ and
	$\varepsilon\sim\Nu_2(0,\varsigma^2I_2)$, the pair is $Z\pm\varepsilon/2$,
	with toroidal wrap-around inside the quadrat; the pair width $\varsigma$ is
	not to be confused with the projection bandwidths $\sigma_b$. Since the
	expected counts per quadrat are unchanged, the signal is invisible at the
	$2\times2$ scale. It is the scale of the pair that decides the scenario: with
	$\varsigma$ well below the mean nearest-neighbour distance under the null,
	$1/(2\sqrt{\lambda_0})=0.029$, the pattern becomes a showcase for
	Clark--Evans and no other procedure comes close to it. We set
	$\varsigma=0.055$ by way of illustration.
		\item The {\emph{Filaments}} scenario keeps the same counts but
	replaces a proportion $\tau$ of the points by filamentary points
	$X=Z+Tv_\vartheta+\eta v_\vartheta^{\perp}$, with
	$v_\vartheta=(\cos\vartheta,\sin\vartheta)$, $Z\sim\Ucal(C_j)$, $T$ the
	longitudinal position and $\eta$ a transversal noise; the filament angle
	$\vartheta$ is a parameter of the alternative, independent of the directions
	$\theta_b$ onto which one projects. Here the departure is purely geometric
	and directional.	
	\item The {\emph{Contrast}} scenario deliberately
	abandons the equality of intensities: one simulates $$N_j\sim\Poi\{\lambda_0w_j(\tau)/4\},$$ with normalized weights
	$w(\tau)\propto\{1+\rho(\tau),1-\rho(\tau),1-\rho(\tau),1+\rho(\tau)\}$,
	$\rho(0)=0$, which keep the total intensity approximately fixed, and the
	filamentary structure is added. The signal is now directly visible in the
	quadrat counts, so that the scenario is a positive control for that
	competitor.	
\end{itemize}
In the three Cox alternatives, $\tau=0$ recovers the homogeneous Poisson
process and larger values strengthen the random heterogeneity of the
intensity.
\begin{itemize}
		\item The {\emph{LGCP bands}} scenario is a log-Gaussian Cox process whose
	intensity exhibits smooth fluctuations organized in bands; it assesses
	sensitivity to heterogeneity of low spatial frequency, for which a projection
	may average away part of the transversal signal.	
	\item The {\emph{Diffuse clusters}} scenario introduces aggregation
	through spatially diffuse clusters, whose diffuseness prevents the signal
	from being reduced to extremely small distances.	
	\item The {\emph{Shot-noise ridges}} scenario is a shot-noise process whose
	random intensity concentrates mass around elongated ridges, combining random
	heterogeneity with directional structure.	
\end{itemize}
\new{For log-Gaussian and shot-noise Cox processes see} \citet{MollerWaagepetersen2003} \new{and}
\citet{DiggleMoragaRowlingsonTaylor2013}\new{;} for Poisson cluster processes\new{, see}
\citet{BaddeleyRubakTurner2015}.
 Figure~\ref{fig:patterns-d2}
displays the departure from the null model as
$\tau$ grows to $1$, in each of the six scenarios.
\subsection{Results}
\label{subsec:sim-results}
\begin{figure}[htbp]
	\centering
	\includegraphics[width=\textwidth]{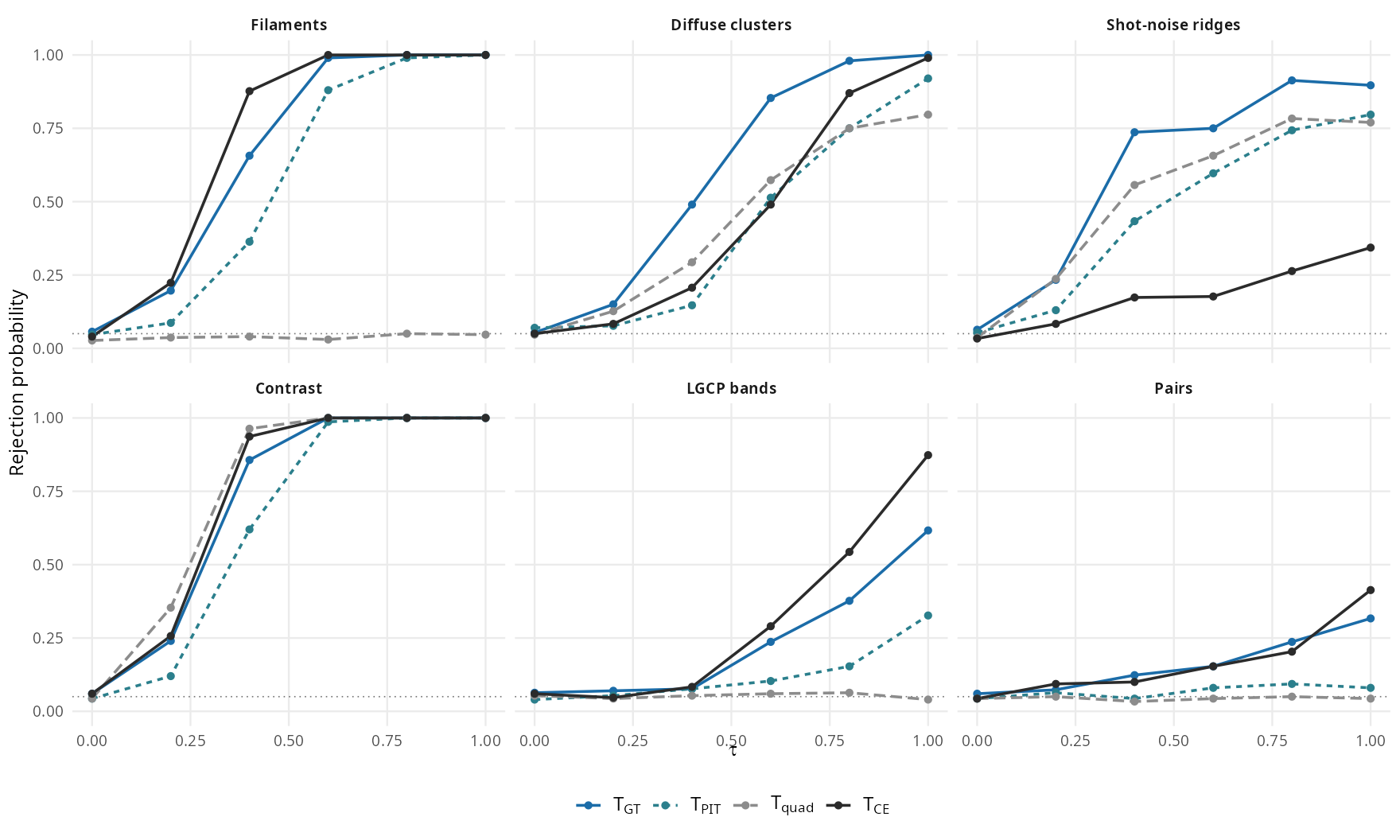}
	\caption{Empirical power functions over the six alternatives with $c_2\ge0$,
		one panel per scenario and four curves per panel; the dotted line marks the
		nominal level $\alpha=0.05$. The three upper panels are the scenarios in
		which random projections are most competitive; the three lower ones include
		the controls that favour the coarse-scale and local competitors}
	\label{fig:power-panel}
\end{figure}

Figure~\ref{fig:power-panel} collects the power functions over the six
alternatives. All figures are rejection frequencies at the nominal level
$\alpha=0.05$.
In {\emph{Diffuse clusters}} and {\emph{Shot-noise ridges}},
$T_\GT$ performs well. The reason is the same in both cases: the signal is of
intermediate spatial scale and is spread over elongated regions, too coarse
for the nearest-neighbour distance to register it and too fine for a
partition into four quadrats to resolve it. This is exactly the gap that the
Gaussian thinning covers, because its bandwidth $\sigma_b$ sweeps precisely
that range of scales.
In {\emph{Filaments}} the scenario preserves the $2\times2$ counts
exactly, so that this panel certifies that the power of $T_\GT$ comes from
the geometry of the pattern and not from a marginal change of intensity{;
here $T_\GT$ is second only to Clark--Evans, and all procedures except
$T_{\mathrm{quad}}$ saturate by $\tau=0.8$}. In
{\emph{Contrast}}, the positive control designed in
favour of the quadrats, these take the lead at $\tau=0.2$ ($0.350$ against
$0.237$){; at $\tau=0.4$ $T_\GT$ is still behind $T_{\mathrm{quad}}$
and $T_{\mathrm{CE}}$, and all three saturate at $\tau=0.6$. In these two
panels the proposed method is competitive without being the most powerful.}
In {\emph{LGCP bands}}, by contrast, the heterogeneity is of low spatial
frequency and the projection integrates mass in the transversal direction{;
$T_\GT$ is second, clearly behind Clark--Evans for large $\tau$}. In
 {\emph{Pairs}} the departure is purely local{; $T_\GT$ tracks
Clark--Evans up to $\tau=0.8$ and falls behind it only at $\tau=1$.}
Figure~\ref{fig:power-panel} also makes it possible to isolate the
contribution of a single ingredient. $T_\GT$ and $T_{\mathrm{PIT}}$ use the
same $B$ directions and the same probability integral transform; the only
thing that separates them is the Gaussian weight. The thinning is thus seen
not to be an ornament of the theoretical construction: it pays in power, and
it pays in every scenario examined. The second comparison is with
$T_{\mathrm{quad}}$, which illustrates the cost of fixing the scale in
advance: it resolves the scenario tailored to it and sits at the nominal
level in three of the six panels.
\subsection{The repulsive case}
\label{subsec:sim-repulsive}
\begin{figure}[htbp]
	\centering
	\includegraphics[width=\textwidth]{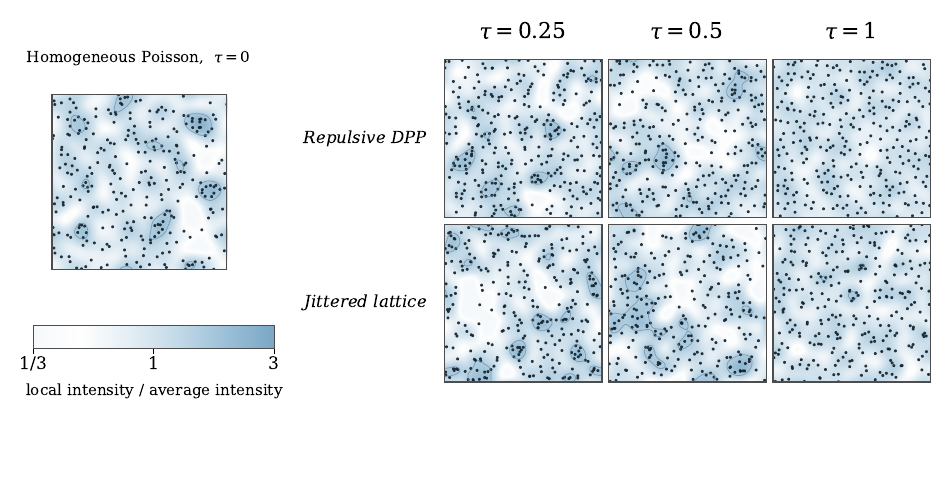}
		\vspace{0.15cm}	
	\includegraphics[width=\textwidth]{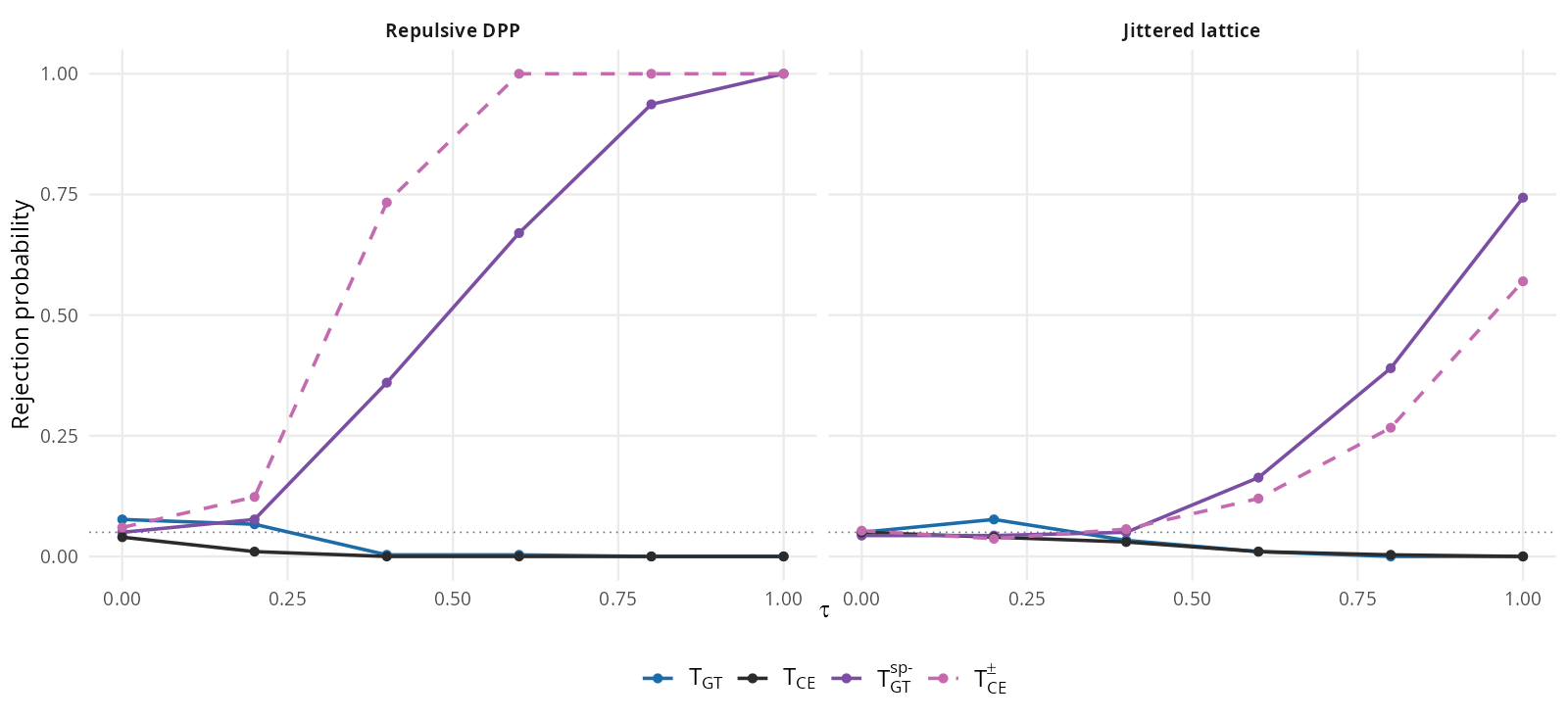}	
	\caption{The repulsive case. Top, simulated patterns at
		$\tau\in\{0,0.25,0.5,1\}$ for the determinantal process and for the
		perturbed lattice; in both the intensity remains fixed and only the
		regularity increases. Bottom, empirical power functions.}
	\label{fig:repulsive}
\end{figure}

Alternatives with $c_2\le0$ are treated separately for three reasons. The
orientation of the statistics changes, because the theoretical departure now
lies in the lower tail; the natural competitor changes, because the one-sided
version of Clark--Evans is blind to regularity by construction; and, above
all, the reading changes: it is here that the choice of discrepancy and of
aggregation rule ceases to be a refinement and becomes what decides whether
the procedure detects anything at all.
The first fact to record is that $T_\GT$ is useless in these scenarios. This
is because $T_\GT$ summarizes the departure over $K_b$ coarse cells, with
about eight expected points per cell. At that granularity the
\emph{under}dispersion produced by regularity is all but invisible: a process
more ordered than Poisson generates cell counts practically
indistinguishable from those of the null model. The information about
regularity lies not in the counts but in the spacings. Moreover, aggregating
by the maximum is the right thing to do when the signal concentrates in a few
directions, as happens with the filaments of
\S\,\ref{subsec:sim-scenarios}. Repulsion, by contrast, is a \emph{coherent}
signal: it pushes all projections slightly in the same direction. Changing
both ingredients, and exploiting the fact that the sign of $c_2$ is known in
advance here, we define
\begin{equation}
	\label{eq:sim-gs-minus}
	T^{\mathrm{sp}-}_{\GT}=-\frac1B\sum_{b=1}^{B}G^{*}_{u_b},
\end{equation}
where $G^{*}_{u_b}$ is the standardized Greenwood spacings score of the
transformed points of projection $u_b$. The Gaussian thinning, the same
projections and the same transformation by the cumulative intensity are all
retained; the only things that change are the univariate discrepancy and the
aggregation rule.
The natural competitor also changes orientation. The one-sided version of
Clark--Evans is blind to regularity by construction, so we use
\begin{equation}
	\label{eq:sim-ce-two-sided}
	T^{\pm}_{\mathrm{CE}}=\left|\log\!\left(\overline R/\overline R_0\right)\right|,
	\qquad
	\overline R_0=\frac{1}{2\sqrt{\widehat\lambda}},
\end{equation}
which in these scenarios coincides numerically with the one-sided version
oriented towards regularity. Four procedures are reported: $T_\GT$,
$T^{\mathrm{sp}-}_{\GT}$, $T_{\mathrm{CE}}$ and $T^{\pm}_{\mathrm{CE}}$.
\paragraph{Two repulsive alternatives.}
The first, {\emph{Repulsive DPP}}, is a homogeneous determinantal point
process $\DPP(C)$ on $\new{A}$ with intensity $\rho=\lambda_0$ and kernel from the
Bessel-type family of
\citet[\S\,5.1]{BiscioLavancier2016}, with $\sigma=0$ and $d=2$,
\begin{equation}
	\label{eq:sim-dpp-kernel}
	C(x)=\rho\,\frac{J_1\!\left(2|x|/\alpha\right)}{|x|/\alpha},
	\qquad
	0\le\alpha\le\alpha_{\max}=\frac{1}{\sqrt{\pi\rho}},
\end{equation}
where $J_1$ is the Bessel function of the first kind. The parametrization
$\alpha=\tau\,\alpha_{\max}$ reproduces the convention of the study: $\tau=0$
is the Poisson process, the non-interacting limiting case of this family, and
$\tau=1$ is $\DPP(C_B)$, the unique kernel that makes the process the most
repulsive, both globally and locally, among all homogeneous determinantal
point processes of intensity $\rho$
\citep[Corollary~3.3]{BiscioLavancier2016}. The simulation uses the
implementation of \citet{LavancierMollerRubak2015} in \texttt{spatstat}
\citep{spatstat}. Here $c_2(x,y)=-C(x-y)^2\le0$ and, as in every
determinantal point process, there is strict exclusion at zero distance:
$g(0)=0$.
The second, {\emph{Jittered lattice}}, isolates the opposite feature. One
starts from a $k\times k$ lattice with $k=\lfloor\sqrt{\lambda_0}\rceil$ and
perturbs each point by a Gaussian displacement with standard deviation $0.55$
times the lattice spacing, with toroidal wrap-around; each perturbed point is
retained with probability $\tau$ and a Poisson process of intensity
$(1-\tau)\lambda_0$ is superposed. Thus $\tau=0$ is exactly the null model,
$\tau=1$ is the perturbed lattice, and the total intensity remains
{approximately $\lambda_0$ (the lattice has $k^2=289$ points for $\lambda_0=300$)}.
\begin{remark}
	\label{rem:dpp-hypotheses}
	A homogeneous determinantal point process does \emph{not} satisfy the Palm
	domination $S\cup\{0\}\preceq S^0$. Determinantal
	point processes are negatively associated \citep{Lyons2003,Lyons2014}, so that they do
	not enjoy the FKG property, and their reduced Palm version is a determinantal
	point process whose kernel is dominated by $C$ in the operator order
	\citep{ShiraiTakahashi2003}; the domination is reversed,
	$S^0\setminus\{0\}\preceq S$, which is the second alternative in
	Theorem~\ref{thm:main-one-projection}(a); see also Lemma~\ref{lem:fkg-palm} and
	Corollary~\ref{cor:fkg-one-projection}\new{, and Remark~\ref{rem:signo-fijo} for an alternative argument}. The remaining hypotheses hold effortlessly: $\kappa(z)=-C(z)^2$ is
	even, continuous and bounded by $\rho^2$.	
	Within the determinantal class, the last link in the identification
	chain also admits a direct argument that does not use Palm domination. If $c_2=0$
	almost everywhere, then $C(z)=0$ for almost every $z\neq0$ and, by continuity
	of the kernel, for every $z\neq0$; hence
	$\rho_n(x_1,\ldots,x_n)=\det[C(x_i-x_j)]_{i,j=1}^{n}=\rho^{\,n}$, which are
	the correlation functions of the Poisson process of intensity $\rho$. Since
	$\rho_n\le\rho^{\,n}$, Lenard's uniqueness condition
	\citep{Lenard1973,DaleyVereJones2003} holds and the correlation functions
	determine the law, so that $S$ is Poisson. Theorem~\ref{thm:main-one-projection}
	therefore holds for determinantal processes also by this route, and  {\emph{Repulsive DPP}} is a legitimate
	alternative for the study.
\end{remark}
\paragraph{Results.}
Figure~\ref{fig:repulsive} shows that the two scenarios are ordered in
opposite ways, and that this inversion is the central finding of this part.
Read together, they indicate that the nearest-neighbour distance summarizes
repulsion in a single local number, which makes it unbeatable when the
repulsion is local and unremarkable otherwise. The projected spacings,
averaged over the $B$ directions, aggregate information from the whole range
of scales that the window can resolve, and for that reason{, in the
\emph{Jittered lattice}, they detect the regularity earlier and more strongly
than the nearest-neighbour distance does.}
This gives operational content to Remark~\ref{rem:dpp-hypotheses} --- the
identification guarantee extends to repulsive processes, and now so does
finite-sample power --- and leaves a concrete recommendation: the aggregation
rule is not an implementation detail but a choice that should match the
expected nature of the alternative. The maximum is appropriate for signals
concentrated in a few directions, such as filaments; the average, for
coherent second-order signals, such as repulsion.
\newpage

\bibliographystyle{abbrvnat}
\bibliography{biblio_poisson,reference,reference1}
\end{document}